\documentclass[useAMS, usenatbib, usegraphicx]{mn2e} 
\usepackage{deluxetable} 

\newcommand{\Trad}{T_{\rm rad}}

\title[Multiphase BH Accretion Flow] 
{Multiphase, non-spherical gas accretion onto a black hole}    

\author[P. Barai et al.] 
{Paramita Barai$^{1, 2}$\thanks{E-mail: pbarai@oats.inaf.it}, 
Daniel Proga$^{3, 1}$, and Kentaro Nagamine$^{1}$\\    
$^{1}$Department of Physics \& Astronomy, University of Nevada, Las Vegas, 4505 S. Maryland Parkway, Box 454002, \\ 
Las Vegas, NV, 89154-4002, USA \\
$^{2}$INAF - Osservatorio Astronomico di Trieste, Via G.B. Tiepolo 11, I-34143 Trieste, Italy \\   
$^{3}$Princeton University Observatory, Peyton Hall, Princeton, NJ 08540, USA}

\begin{document} 



\maketitle

\label{firstpage}

\begin{abstract}

We investigate non-spherical behavior of gas accreting onto a central supermassive black hole.   
Assuming optically thin conditions, we include radiative cooling and radiative heating by the central X-ray source.   
Our simulations are performed using the 3D SPH code GADGET-3 
and are compared to theoretical predictions as well as to 1D simulations performed using the grid code ZEUS.   
As found in earlier 1D studies, our 3D simulations show that the accretion mode depends on    
the X-ray luminosity ($L_X$) for a fixed density at infinity and accretion efficiency.  
In the low $L_X$ limit, gas accretes in a stable, spherically symmetric fashion.   
In the high $L_X$ limit, the inner gas is significantly heated up and expands, reducing the central mass inflow rate.   
The expanding gas can turn into a strong enough outflow capable of expelling most of the gas at larger radii.   
For some intermediate $L_X$, the accretion flow becomes unstable developing prominent non-spherical features.  
Our detailed analysis and tests show that the key reason for   
this unstable non-spherical nature of the flow is thermal instability (TI).   
Small perturbations of the initially spherically symmetric accretion flow that is heated by the intermediate $L_X$    
quickly grow to form cold and dense clumps surrounded by overheated low density regions.   
The cold clumps continue their inward motion forming filamentary structures;    
while the hot infalling gas slows down because of buoyancy and   
can even start outflowing through the channels in between the filaments.    
We measured various local and global properties of our solutions.    
In particular, we found that the ratio between the mass inflow rates of the cold and hot gas is a    
dynamical quantity depending on several factors:   
time, spatial location, and $L_X$; 
and ranges between $0$ and $4$.   
We briefly discuss astrophysical implications of such TI-driven fragmentation of accreting gas    
on the formation of clouds in narrow and broad line regions of AGN,  the formation of stars,   
and the observed variability of the AGN luminiosity.     

\end{abstract}


\section{Introduction} 
\label{sec-intro}  

Active galaxies are believed to be powered by accretion of matter onto the  
central supermassive black holes (SMBHs) \citep[e.g.,][]{Salpeter64, LyndenBell69, Blandford76,  
Ozernoi78, Balick82, Rees84, Fabian90, Kormendy95, Kauffmann00, Ferrarese05}.   
Active Galactic Nuclei (AGN) are argued to influence cosmological galaxy formation in the form of feedback,  
whereby the overall properties of a galaxy can be regulated by its central BH   
\citep[e.g.,][]{Silk98, King03, Wyithe03, Granato04, Murray05, Begelman05, Best07, Ciotti07, Pipino09, Ostriker10, DiMatteo11}.   
Observationally AGN are characterized as a compact continuum source around a central SMBH,  
surrounded by a much larger emission line region consisting of multiphase gas clouds.   
The continuum emission shows a    
relatively flat spectral energy distribution   
over a broad wave-band from radio to gamma rays, and is modeled as coming from the   
central engine composed of a gas accretion disk of size $10^{-4} - 10^{-3}$ pc.   
The broad-line region (BLR) extends up to $0.01 - 1$ pc and is composed of higher-density, higher-velocity gas;   
while the outer narrow-line regions (NLR) are considered to be located at distances up to $10 - 1000$ pc from the central SMBH. 

Modeling gas accretion onto SMBHs and resulting AGN feedback in a cosmological context is   
computationally challenging, because a large dynamic range of length scales is involved:   
BH accretion on sub-pc scale, to galaxy physics on hundreds of kpc scale.   
Therefore SMBH accretion and feedback is usually treated by a sub-grid model in cosmological simulations   
\citep[e.g.,][]{DiMatteo08, Booth09}.   
At the same time, gas accretion flow within the Bondi radius of a SMBH   
has just started to be resolved in observations \citep{Wong11}.   
Hence numerical studies resolving the Bondi radius are required for a meaningful comparison,   
and it would be       
useful to improve a sub-grid model of AGN feedback for cosmological simulations based on the results of small-scale simulations. 
In \citet[][hereafter Paper~I]{Barai11}, we studied  
spherically-symmetric Bondi accretion of gas onto a SMBH,     
and started to explore the effects of radiative heating and cooling.   


A number of recent studies resolving the Bondi radius have focused on accretion onto intermediate-mass BHs   
regulated by feedback from X-ray and UV radiation emitted isotropically near the BH.   
Analytical work \citep{Milosavljevic09b}   
as well as 1D and 2D simulations \citep{Milosavljevic09, Park11, Park12}   
find that photoionization heating and radiation pressure significantly reduce the steady-state accretion rate   
to a fraction of the Eddington rate, or over 2 orders of magnitude below the Bondi rate.  
However, in 1D simulations by \citet{Li11}   
the self-gravity of the gas eventually overcomes the radiative feedback effects,  
and boosts the accretion to the Eddington rate after one free-fall timescale.   
\citet{Park11, Park12} also find the development of Rayleigh-Taylor instabilities in their 2D simulations.   
In these work the BH luminosity is made to be proportional to the rate at which gas accretes     
across the inner boundary, whereas in our study the central luminosity has a constant value in each simulation    
because we want to have a simplified setup with a non-variable source   
to enable us cleanly investigate the physical processes influencing the accretion flow.   

In Paper~I we discovered indications of thermal instability (TI) in some runs,   
which we investigate in detail in the current work.  
Studying TI in a dynamical flow in 3D is challenging, because the effects of a gravitational field   
significantly complicate the development of TI \citep[e.g.,][]{Balbus89}.    

The foundations for understanding astrophysical thermal instability were laid in the classical paper by \citet{Field65}.  
The general criterion for TI to occur is: 
\begin{equation} 
\label{eq-TI-criterion} 
\left [ \frac{\partial {\cal{L}}} {\partial S} \right ]_A < 0,
\end{equation} 
where ${\cal{L}}$ is the net cooling-heating rate, $S$ is the specific entropy,  
and $A$ is a thermodynamic variable that is kept constant for a given perturbation.  
Subsequently the theory of TI has been studied over decades, including analytical developments:  
local TI in static and dynamical systems \citep[e.g.,][]{Balbus86, Balbus89, Balbus95},  
global TI by linear perturbation analysis \citep[e.g.,][]{Yamada01}, 
and numerical hydrodynamical calculations to investigate the perturbations in the non-linear regime \citep[e.g.,][]{Hattori90}.   


TI has been invoked to explain observed features in various astrophysical domains:  
solar prominences \citep[e.g.,][]{Parker53, Karpen88}; 
formation of cold gas clouds and clumpy substructure in the interstellar medium   
\citep[e.g.,][]{Schwarz72, Parravano87, Hennebelle99, Burkert00, Koyama02, Sanchez-Salcedo02, Inutsuka05};  
origin of globular clusters during the collapse of a protogalaxy \citep[e.g.,][]{Fall85};  
formation of galaxies \citep[e.g.,][]{Gold59, Sofue69};  
general cooling flows in galaxies and galaxy cluster cores  
\citep[e.g.,][]{Mathews78, Nulsen86, Malagoli87, Meiksin88, Yoshida91, Guo08};  
the multi-phase structure of cold filaments condensing out of the hot intracluster medium  
in galaxy clusters and groups \citep[e.g.,][]{Sharma10, McCourt11}.   

More relevant for our work are the applications of TI in black hole accretion systems.  
Spherically symmetric accretion flow onto a compact object is preheated by radiation from the central X-ray source.  
\citet{Ostriker76} showed the preheating suppresses accretion,  
such that above a X-ray luminosity $0.005$ of the Eddington limit steady flow is impossible.  
\citet{Cowie78} found time-dependent behavior of the preheated accretion  
for a much wider range of source parameters, and the presence of instabilities.  
However, \citet{BisnovatyiKogan80} found that numerical solutions exist for any luminosity below the Eddington limit,  
but they did not investigate the stability of the solutions in detail.  
Performing stability analyses \citet{Stellingwerf82} found globally stable solutions, but those become subject to  
a drift instability at luminosities over the Eddington limit exceeding $\sim 0.03$.    
Hydrodynamical simulations by \citet{Proga07} show that with an accretion luminosity of $0.6$ of the Eddington value,  
the gas flow settles into a steady state with two components:   
an equatorial inflow, and a bipolar inflow/outflow with the outflow leaving the system along the disk rotational axis.   
There the temperature at the outer radius was fixed at a relatively high value ($2 \times 10^{7}$ K),  
so the gas did not go through a thermally unstable phase.  
Also \citet{Park12} find that a quasi-steady BH accretion is possible at the Eddington luminosity   
when the gas density is above a critical value inversely proportional to the BH mass.  

Investigating spherical accretion of gas irradiated by a quasar-like continuum,  
\citet{Krolik83} found specific criteria for when TI may disrupt single-phase steady flow,   
in the luminosity -- efficiency (of mass to energy conversion) plane.  
\citet{Krolik88} argued that broad emission line clouds in AGN originates in cooler condensations  
forming within a hot medium via finite-amplitude TI.    
However, \citet{Mathews90} concluded that broad line-emitting clouds    
are unlikely to arise from thermally unstable condensations in a hotter medium,    
but are likely composed of higher-density gas ejected from nearby stars or from a dense accretion disk.   
\citet{Wang12} investigated the dynamics of clumps, considered to exist by thermal instability,   
embedded in and confined by advection-dominated accretion flows onto black holes.   


We find that a spherically accreting gas distribution acted upon by a central gravitational potential,  
undergoing radiative heating and cooling, can be thermally unstable, cool non-spherically and become clumpy.  
Hence results from our simulations show that  
fragmentation can occur in a simple system with minimal physical processes.  
Our study can give insight into the formation and evolution of clouds      
near AGN.  
TI operating in such a way can be responsible for the origin of two phase medium in the NLR,  
support star-formation near AGN,  
contribute to the observed variability of the luminosity.     

This paper is organized as follows. We describe our simulations in \S\ref{sec-numerical}.  
We present and explain the results in \S\ref{sec-results}, and discuss in \S\ref{sec-discussion}.  
We summarize and conclude in \S\ref{sec-conclusion}.


\section{Simulations} 
\label{sec-numerical}   

In Paper~I we performed simulations of gas accretion onto a SMBH of mass $10^8 M_{\odot}$  
within the radial range of $0.1 - 200$ pc, using the 3D SPH code GADGET-3 \citep[originally described by][]{Springel05}. 
The gas experiences gravitational potential of the central SMBH only, and there is no self-gravity.  
We first tested how accurately the GADGET code can follow the Bondi problem,  
aiming to know what was required for it to reproduce the Bondi solution.  
Our simulations could reproduce adiabatic Bondi accretion for durations of  
a few dynamical times at the Bondi radius, and for longer times if the outer radius is increased.  
Next we investigated one of the modes of AGN feedback: radiation heating, by performing simulations of  
gas accretion without and with radiative heating by a central X-ray corona and radiative cooling.  
We found that artificial viscosity causes excessive unphysical heating near the inner radius.  
We also observed feedback due to radiative heating in our simulations,  
as increasing the X-ray luminosity ($L_X$) produced a lower central mass inflow rate,  
and an outflow developed for sufficiently high $L_X$.  

All the results of gas properties presented in Paper~I were spherically-symmetric. 
However, for certain parameters, we observe that the spherical-symmetry is broken,  
and the flow becomes inhomogeneous and multi-phase.  
Here, we analyze these non-spherical simulations and present corresponding results.   
We investigate the gas behavior in detail, aiming to comprehend the underlying physics versus numerics at play.  

The radiative heating and cooling prescriptions used in our simulations are described in \S2.4 of Paper~I.  
Radiation from the central SMBH is considered to be in the form of a  
spherical X-ray emitting corona of luminosity $L_X$ which irradiates the accretion flow,  
and we consider an optically thin case.  
The local X-ray radiation flux at a distance $r$ from the central source is $F_X = \frac{L_X}{4 \pi r^2}$.  
The photoionization parameter $\xi$ is defined as  
\begin{equation}  
\label{eq-smallXi}  
\xi \equiv \frac{4 \pi F_X}{n} = \frac{L_X}{r^2 n}.   
\end{equation}  
Here $n = \rho / (\mu m_p)$ is the local number density of gas, computed using the mass density $\rho$,  
mean molecular weight $\mu$, and the proton mass $m_p$.  
The heating-cooling function is adopted from \citet{PSK00}, originally from \citet{Blondin94}.  
These are approximate analytic formulae for the heating and cooling rates of an X-ray irradiated   
optically-thin, cosmic abundance (in this context meaning close to solar metallicity)    
gas illuminated by a $10$ keV bremsstrahlung spectrum.   
Radiation pressure of the gas has not been included.  
The net heating-cooling rate, ${\cal{L}}$, is given by   
\begin{equation}
\label{eq-netL}
\rho {\cal{L}} = n^2 \left( G_{\rm Compton} + G_{X} - L_{b,l} \right) = {\rm Function~of~}  \left(\xi, T \right),   
\end{equation}  
which includes Compton heating and cooling,  
X-ray photoionization heating and recombination cooling, and bremsstrahlung and metal line cooling.  
The characteristic temperature of the bremsstrahlung radiation has a constant value $T_X = 1.16 \times 10^{8}$ K.  
The heating-cooling rate depends on density, since $\xi$ is a function of $\rho$ and $r$.  

We compute an additional quantity for the analysis presented in this work:  
the pressure-ionization parameter $\Xi$ \citep[e.g.,][]{Krolik81}, which is defined as
\begin{equation} 
\label{eq-Xi}  
\Xi \equiv \frac{F_X}{c P} = \frac{L_X}{4 \pi r^2 c P},  
\end{equation}  
where $c$ is the speed of light, and $P$ is the gas pressure.   

The functional form of radiative heating and cooling in Equation~(\ref{eq-netL})  
has been implemented in the grid code ZEUS \citep[e.g.,][]{Proga07, KP09a, KP09b},  
and there are many resulting insights from ZEUS simulations about the dynamics of accreting gas irradiated by a quasar.   
Considering radiation field from a UV-emitting standard accretion disk  
along with a spherical central X-ray source and line driving,  
\citet{Proga07} found that the accretion flow settles into a steady state and has two components (\S\ref{sec-intro}).  
\citet{POK08} added gas rotation (also in \citealt{KP09a, KP09b}), and saw   
an increase of the outward thermal energy flux, leading to fragmentation and time variability of the outflow.    

As a new feature, in Paper~I we increased the outer radius to $r_{\rm out} = 200$ pc. 
This in turn allows us to follow the evolution of gas that is cooler than in \citet{KP09b} ($r_{\rm out} \sim 10$ pc),  
so that the conditions are met for the gas to undergo thermal instability.  
The gas near $r_{\rm out}$ is subsonic and    
lies in the cool, thermally stable branch of the $T - \xi$ radiative equilibrium curve (\S\ref{sec-TI}).   
As it accretes inward it goes to the thermally unstable branch.   

Table~\ref{Table-AccrModes} gives a list of four simulations presented here. 
The run numbers correspond to those in Paper~I. 
The only parameter varied between the runs is the X-ray luminosity $L_X$, in units of the Eddington luminosity $L_{\rm Edd}$. 
These simulations are done by restarting the Run 23 (presented in Paper~I, with $L_{X} = 5 \times 10^{-4}$) at $t = 1.4$ Myr, and performing subsequent Runs $25 - 28$ with $L_{X} = 0.005, 0.01, 0.02, 0.05$.  
Some of the common parameter values are: 
$\gamma_{\rm run} = 5/3$, $r_{\rm in} = 0.1$ pc, $r_{\rm out} = 200$ pc, 
$\rho_{\infty} = 10^{-23}$ g/cm$^3$, and $T_{\infty} = 10^5$ K.  
The temperature at the outer radius corresponds to a Bondi radius of $R_B = 183.9$\,pc.  
The Bondi accretion rate is $\dot{M}_{B} (\gamma = 5/3, \rho_{\infty}, T_{\infty}) = 4.9 \times 10^{25}$ g/s.  
This is $0.35$ times the Eddington accretion rate, assuming a radiative efficiency of $0.1$.  

An assumption of our work is a constant value of $L_X$ in a simulation.  
This enables us to have a simple setup in order to  
investigate the physical origin of the non-spherical behavior of the accretion flow.   
A more realistic scenario would be $L_X$ as a function of the mass accretion rate at the inner boundary.   
However this would lead to an additional dependence, which we leave for future studies.

\section{Results}  
\label{sec-results} 

\subsection{$L_X = 0.01$: Cold Clumps}         
\label{sec-LX-0.01} 

There is fragmentation and clumping in Run 26, as the gas cools and becomes denser in certain regions.   
The free-fall time at the outer radius is $2.98$ Myr.  
Non-spherical features start to develop from $\sim 1.6$ Myr continuing as long as   
the simulation is run ($2.12$ Myr), occurring at $r < 30$ pc.  

Figure~\ref{fig-Splash-Lx0.01-30pcYZ} shows the inner $30$ pc cross-section  
in the $[x - y]$ plane through $z = 0$ at $t = 2.047$ Myr,  
with the gas density in the top-left, temperature in the top-right, photoionization parameter in the bottom-left,  
and Mach number in the bottom-right panels, overplotted with velocity vector arrows   
(the scale is indicated at the top-left in units of km/s).   
The figure illustrates the filamentary structure of the cold inflow.  
The cold-dense filaments at $r > 5$ pc have a lower-photoionization, and extend up to $r \sim 20 - 30$ pc.   
Hotter, lighter gas exists in-between the filaments.  
Both the hot and cold components flow inward initially.      
The hot phase is subsonic, while  
the cold phase is supersonic reaching Mach $\sim 10$ all the way down to $r_{\rm in}$.  
At $t > 1.8$ Myr some of the hotter gas between $r \sim 25 - 30$ pc start to flow outward,   
but there is no well-developed outflow, since the hot gas encounters colder inflowing gas at larger radii.  
The photoionization parameter is high near the center despite the high central density.   

The details of the inner-most cold clumps are plotted in Figure~\ref{fig-Splash-Lx0.01-4pcXY}.  
It is a cross-section of the inner $4$ pc of the $[x - y]$ plane at $t = 2.047$ Myr  
(same plane as Figure~\ref{fig-Splash-Lx0.01-30pcYZ}, zoomed-in here).  
However note that this cross-section uses a different color scheme and it has been plotted without the velocity vectors,  
in order to show the small-scale features clearly.  
The central $r < 1$ pc region is denser than the surrounding,   
since the stable free-fall gas density increases going inward (e.g., Figure~7 in Paper~I).   
In terms of other properties this central region is characterized by high temperature,   
high photoionization parameter, and supersonic velocity.   

We investigate the motion of the outer clumps as they fall into $r_{\rm in}$ by making  
movies\footnote{http://adlibitum.oats.inaf.it/barai/AllPages/visualization.html}  
of multiple snapshots within $30$, $12$ and $4$ pc cross-sections.  
After formation from thermally-unstable gas (details in \S\ref{sec-TI}), as the clumps fall in        
they undergo a dynamic evolution.  
Some stretch and fragment producing two or more sub-clumps,    
caused by the gravitational field of the central SMBH.    
Sometimes smaller clumps merge with each other creating a bigger one.  

In the inner volume, at $r < 0.5$ pc,   
although the density of the falling clumps remains significantly higher than the mean density,   
the temperature becomes comparable to the mean value at a given radius.   
Consequently we can see a dense clump in the density-plane, but it is not cold any more in the temperature-plane.   
An example in Figure~\ref{fig-Splash-Lx0.01-4pcXY} is the distinct dense blob at $y \sim -0.5$ pc,   
which is also visible as a low-photoionization and high-Mach clump.   
The central heating at $r < 1$ pc is mostly by adiabatic compression where the entropy of a gas particle remains constant.   
Radiative heating would dominate at a few pc for some particles.    
We have checked the role of artificial viscosity (given our results in Paper~I),   
and found that although it increases in magnitude toward $r_{\rm in}$,   
it always remains sub-dominant compared to adiabatic heating in the simulations presented here.   

The gas particle properties as a function of radius at $t = 2.0$ Myr is plotted in Figure~\ref{fig-Scatter-Lx0.01}.   
It can be compared to Figure~15 in Paper~I which shows the same simulation run   
at an earlier time, $t = 1.5$ Myr, where the flow is still stable and spherical.   

Figure~\ref{fig-Scatter-Lx0.01} shows radial scatters at $r < 30$ pc as large as 2$-$3 orders of magnitude. 
It corresponds to the multi-phase structure of the gas,   
with a group of particles in the hot branch ($T > 10^6$ K), another group in the cold branch ($T < 3 \times 10^4$ K),  
and some in transition between the two.    
The cold particles are adiabatically heated up at $r < 0.7 - 1$ pc, with no particles below $10^5$ K at those inner radii.   
The temperature increases to $> 10^7$ K near $r_{\rm in}$.   
Some particles are in free-fall, some are slower forming a scatter by a factor of a few, 
and some outflowing ($v_r > 0$) between $7 - 25$ pc.    
In the temperature versus ionization parameter planes ($T - \xi$ and $T - \Xi$), the particles are    
scattered around the respective unstable branch (see \S\ref{sec-TI}) of the radiative equilibrium curve.   
The group of particles        
departing from the radiative equilibrium curve with little scatter in the lower-right portion of these planes     
corresponds to the outflow near $r_{\rm out}$.

\subsubsection{Thermal Instability}   
\label{sec-TI}    

The slope of a perturbation in the $[T - \xi]$ and $[T - \Xi]$ planes can be used to determine    
if the gas is stable against such a perturbation.   
The relations for three common processes are: 
\begin{eqnarray}
T \propto \xi^2,  T \propto \Xi^{-2} ({\rm adiabatic~free~fall, ~or,~in~Eq.~(\ref{eq-TI-criterion})} ~ A = S), \\
T \propto \xi, ~ \Xi = {\rm constant} ({\rm constant~pressure, or,} ~ A = p), \\
\xi = {\rm constant},~ T \propto \Xi^{-1} ({\rm constant~density, or,} ~ A = \rho). 
\end{eqnarray} 
These relationships are shown as the blue (dotted, dashed, and dash-dotted) lines  
in the bottom row of Figure~\ref{fig-Evolution-Single}.   

According to our radiative heating and cooling prescriptions, 
the upper-left half of the $\Trad - \xi$ equilibrium curve  
(red line in the bottom-left panel of Figures~\ref{fig-Scatter-Lx0.01} and \ref{fig-Evolution-Single})  
is the net cooling zone, and the lower-right half is the net heating zone.  
Let us consider a perturbation with constant pressure on a point on the $\Trad - \xi$ equilibrium curve.  
The accretion flow is stable in the parameter region where the instantaneous slope of the equilibrium curve   
is shallower than $T\propto \xi$, and vice versa.   
The $\Trad - \xi$ equilibrium curve has the steepest slope between $\xi \sim 100 - 500$,   
therefore it is easier to find a perturbation with a shallower slope than the equilibrium curve,       
making it most prone to instability there.  We call this region as the {\it unstable} region. 
In our simulations we indeed find that gas particles undergo TI within the unstable $\xi$-range, as described below.  

We also compute $T - \xi$ solution of the isobaric instability criterion given in the Equation~(5) of \citet{Balbus86} for local TI in dynamical systems.  
Using $\xi \propto 1 / \rho$, the relevant expression becomes
$$\left( \frac{\partial \ln {\cal{L}}} {\partial \ln T} \right)_{\xi} + \left( \frac{\partial \ln {\cal{L}}}{\partial \ln \xi} \right)_{T} - 1 = 0,$$
which gives the green line in the bottom-left panel of Figure~\ref{fig-Evolution-Single}.  
It exactly overlaps with the $\Trad - \xi$ radiative equilibrium curve in the unstable region.   

By selecting a single particle we study the time evolution of its various radial properties,    
and its thermal history in the $[T - \xi]$ and $[T - \Xi]$ planes.   
Using this approach for particles selected from different parts of the $[T - \xi]$ plane 
(bottom-left panel in Figure~\ref{fig-Scatter-Lx0.01}),  
we find that at initial times ($t < 1.6$ Myr, when the flow is still stable at all-$r$)   
every particle has an increase in temperature while moving inward.   
After $1.6$ Myr we find two broad categories of evolution.   
In the first kind, a particle cools down to $T \sim 3 \times 10^4$ K,   
might undergo alternative periods of heating and cooling while continuing to move in,   
and finally heats up near $r_{\rm in}$.  
These cooled particles form the dense filaments seen in Figures~\ref{fig-Splash-Lx0.01-30pcYZ}, \ref{fig-Splash-Lx0.01-4pcXY}.  
In the second kind, a particle would never cool, but continue to heat up all the way to the center.  
These hot particles form the lighter gas in-between dense filaments.   

We argue that this cooling (which occurs in a non spherically-symmetric manner) is caused  
by the TI in accreting gas acted upon by the adiabatic and radiative processes.   
Investigating the thermal history,  
we find that a single particle goes through both adiabatic and radiative process, each of which dominating 
at different times.  We hereafter call these epochs as {\it adiabatic} and  {\it radiative regimes}. 
In the {\it radiative regime} the particle follows the radiative equilibrium temperature $(\Trad)$,  
and in the {\it adiabatic regime} it is heated up following the adiabatic law $(T_{\rm ff, a} \propto {r} ^{-1})$.   
We determine the dominant process by comparing the terms in the $du / dt$ (time derivative of the specific internal energy) equation, as well as comparing the particle evolution to the free-fall slopes in the $[T - r]$ and $[\rho - r]$ planes.  
We also verify that its entropy remains constant when adiabatic processes dominate.   

We note that the adiabatic free-fall relation, $T_{\rm ff, a} \propto \xi_{\rm ff, a}^2$,  
has the same slope as the radiative equilibrium curve in the unstable region.    
Also in the $[T - \Xi]$ plane, the adiabatic free-fall relation ($T_{\rm ff, a} \propto \Xi_{\rm ff, a}^{-2}$)  
has the same slope as $\Trad$ in the unstable region, where $dT / d\Xi < 0$ (isobaric case).  
This can be seen in the bottom two panels of Figure~\ref{fig-Evolution-Single} as the overlapping red and blue-dotted lines.  

During the linear phase of TI growth, a subsonic particle lying on an unstable part of the   
$\Trad - \xi$ equilibrium curve would move along the constant-pressure ($T \propto \xi$) slope.   
However our system is very dynamical and the non-linear development of TI can be different from classical descriptions.  
Therefore it might not end up on another stable equilibrium branch.   
The adiabatic heating (or, cooling) arising from inward (or, outward) gas motion can prevent   
a particle from following a path with constant-pressure all the way to the next stable position.    


We detect a distinct mode of cooling in the $[T - \xi]$ plane, which occurs as below.   
A particle while being in the adiabatic regime moves up to higher-$T$ along the unstable part of $\Trad - \xi$.   
This happens because of the equality of the adiabatic and radiative slopes in the unstable regions, as described before.    
At a later time the particle has a transition from adiabatic to radiative regime.  
Then TI operates, and the particle undergoes catastrophic cooling or heating (depending on which is dominant),  
and moves away from the unstable $\Trad - \xi$ part.  
It finally settles on one of the stable branches where $\Trad - \xi$ has a smaller slope.     

Figure~\ref{fig-Evolution-Single} plots the evolution of a typical particle which goes through such a cooling instability,  and demonstrate the above arguments.  
This particle undergoes evolution of the first kind: heats up initially (A$\rightarrow$B$\rightarrow$C), cools (C to D), and finally heats up again near the center (D$\rightarrow$E$\rightarrow$F$\rightarrow$G).  
In between there could be several episodes of alternative heating and cooling each lasting for short durations,   
as we describe below.  

Let us follow the particle evolution on the $\Trad - \xi$ plane in more detail. 
Starting from the IC at $t = 1.4$ Myr when it is at point A, $r = 53$ pc indicated by the triangle in each panel,  
the "$+$" symbols denote uniform time intervals of $0.004$ Myr  
(except the two inner-most points F and G which are separated by $\sim 0.001$ Myr).  
The ending point G is at $t = 1.8$ Myr and $r = 0.99$ pc denoted by the square symbol in each panel,  
after which the particle is accreted within $< 0.001$ Myr.  
The IC has been taken from a run with a lower-$L_X$, but with the same density, which corresponded to a lower-$\xi$.   
On starting this run with a higher-$L_X$ the particle has a higher-$\xi$.   
Therefore it is radiatively heated up to $\Trad$ initially from point A to B,  
a phenomenon that happens for all the particles in the simulation.     

After this initial quick heating,   
the particle reaches the point B belonging to the unstable branch of $\Trad$.   
However by that time it enters the adiabatic regime at $r \sim 50$\,pc, 
and moves up along the unstable $\Trad - \xi$ branch from point B to C due to adiabatic heating. 
This can be verified in the $[T - r]$ and $[\rho - r]$ planes where it follows the free-fall slope,  
and its entropy remains constant.  
At point C, $r \sim 35$ pc, it transits to the radiative regime with dominance of cooling and TI operates.   
It cools down rapidly from C to D deviating slightly to the left of equilibrium curve (deviation is more evident in the $T-\Xi$ plot), i.e. in the cooling zone. 
By point D, $r \sim 15$ pc, it reaches the cold-stable $\Trad$.  
Then it becomes dominated by radiative heating for some time, and  
at $r \sim 10$ pc undergoes another transition to adiabatic regime, all of which happens between points D and E. 
It reverts back and radiatively cools between $r = 7 - 5$ pc reaching point E.  
After that it heats up again while falling inward, first from E to F by radiative heating.  
The final heating from F to G at $r < 2$ pc is adiabatic, caused by the central compression of the gas.   
The transition of E$\rightarrow$F$\rightarrow$G can also be regarded as a TI, since it rapidly deviates from the stable branch of the equilibrium curve and landing on another stable branch. 


\subsubsection{Timescale Comparison}   
\label{sec-TI-GrowthTime}    

An important factor for TI to develop is the gas time scales,  
as has been discussed by other authors \citep[e.g.,][]{Beltrametti81, David88, Mathews90}.   
If unstable perturbations occur in supersonic gas, then there are no resulting dynamical changes,   
since pressure forces play only a minor role \citep[e.g.,][]{Krolik83}.   
In other words, the gas with thermal properties in the TI zone should be subsonic,  
otherwise there will not be enough time for TI to grow.   
For example this is the reason that simulations with $L_X < 0.01$ are smooth and spherically-symmetric.  

We investigate the origin of gas clumping in this run   
by comparing the relevant timescales.  
We compute the TI growth time $(t_{gr})$ using the growth rate $(n_1)$ from Equation~(31) of \citet{Field65},   
which in our notation is,  
\begin{equation}  
\label{eq-tGrowth}   
t_{gr} = \frac{1} {n_1} = \left[ \frac{(\gamma - 1) \mu} {\gamma k T_0}   
\left\{ T_0 \left( \frac{\partial{\cal{L}}} {\partial T} \right)_\rho -  
\rho_0 \left( \frac{\partial{\cal{L}}} {\partial \rho} \right)_T \right\} \right]^{-1}.  
\end{equation}   
Here $\rho_0, T_0$ are the unperturbed density and temperature; where we use the current gas properties   
at a given radius when the flow is still spherically-symmetric.    
We analyze the ratio of $t_{gr}$ with the free-fall timescale, $t_{ff} = \left( \frac{r^3} {2 G  M_{BH}} \right)^{1/2}$,   
which is plotted in Figure~\ref{fig-timeGrowthTI-Mach} top panel at three simulation epochs.   
We find positive $n_1$ in some radial ranges representing regions where TI can grow, and    
negative $n_1$ elsewhere    
indicating regions stable to TI seen as the radial gaps in Figure~\ref{fig-timeGrowthTI-Mach}.   

We find that within the positive $n_1$ regions,    
$t_{gr}$ starts to become less than $t_{ff}$ inside a range $r \sim 10 - 80$ pc     
soon after start of the simulation at $t = 1.4$ Myr.    
But the gas do not clump as soon and over that entire $r$-range because it is still supersonic,   
seen in the bottom panel of Figure~\ref{fig-timeGrowthTI-Mach} where Mach number is plotted.    
As discussed at the beginning of \S\ref{sec-LX-0.01},    
non-spherical features first appear from $\sim 1.6$ Myr, at $10 - 20$ pc.    
This is when and where the gas becomes subsonic is addition to $t_{gr} < t_{ff}$,   
from Figure~\ref{fig-timeGrowthTI-Mach}, and therefore clumping develops, as expected.    
\citet{Krolik83} showed that unstable accretion flow, which is subsonic,   
can be disrupted when $t_{gr}/t_{ff}$ becomes small, exactly as we see in our simulations.    


As a quantitative analysis of the growth of instability,  
we plot the range of perturbation amplitude in Figure~\ref{fig-Growth-Perturbation}.  
It shows the time evolution of the temperature and density perturbation at $r = 10$\,pc  
within a width of $\Delta r = 1$\,pc. 
Tracking the gas particles lying within that radial bin,  we plot the minimum, average, and maximum
values as a function of time.  
The corresponding average value at $t = 1.42$ Myr is considered as the unperturbed gas property.  
This time is close to the initial time of $1.4$ Myr  
but distant enough such that the gas has adjusted to the effect of higher-$L_X$ at the beginning,  
and when the flow is still stable and spherical.  
We plot the 
statistics relative to the initial, unperturbed value at $t = 1.42$ Myr. 

The minimum, maximum and average values are all close to the unperturbed average 
at $t= 1.42 - 1.55$ Myr, and they start to deviate after that.  
The perturbations show an exponential growth at first,    
and then come to a saturation for the remaining of the simulation.  
There is an exponential decrease of $T_{\rm min}$ between $1.55 - 1.65$ Myr reaching a value $0.02$,  
indicating the growth of cooling instability, and $T_{\rm min}$ remains almost constant after that.  
A gradual increase is seen for $T_{\rm max}$ from $1.55$ Myr to $2.17$ Myr.  
The value of $\langle T \rangle$ decreases to $0.6$ at $t=2.17$ Myr.  
The value of $\rho_{\rm max}$ exponentially increases to 200 between $t=1.55 - 1.65$\,Myr, 
indicating the growth of cooled, dense clumps due to TI.  
It subsequently fluctuates between $100 - 200$.  
Following a similar trend, $\langle \rho \rangle$ exponentially increases to $10$ by $t=1.7$\,Myr,   
and remains constant after that.  
The value of $\rho_{\rm min}$ gradually decreases to $0.4$ at $t=2.17$ Myr.

\subsubsection{Fraction of Cold vs. Hot Gas in Accretion Flow}   
\label{sec-Cold-Hot-Frac}    

In our simulations we have cold and hot phases of gas forming self-consistently in the accretion flow  
acted upon by radiative heating and cooling (whose functional form are in Paper~I, Section~2.4) and undergoing TI.  
Furthermore we resolve the Bondi radius.  
The multi-phase gas has a temperature varying from $2 \times 10^4$\,K to $10^6 - 10^7$ K\,at the same $r$ in Run 26.   
We choose a limiting temperature of $T_{\rm lim} = 10^5$ K to separate the cold and hot phases. 
We measure the mass inflow rate of cold and hot gas ($\dot{M}_{\rm in, cold}$ and $\dot{M}_{\rm in, hot}$) 
by counting the particles with $v_r < 0$, below and above $T_{\rm lim}$.   
We define the ratio of mass inflow rates of cold gas over hot:  
$\chi = \dot{M}_{\rm in, cold} / \dot{M}_{\rm in, hot}$.  
It is similar to the $\alpha$-parameter multiplied to the Bondi accretion rate  
in some of the galaxy simulations (discussed in \S~\ref{sec-discussion}), which accounts for the fraction of cold, dense gas that is missed due to lack of resolution. 

We find that $\chi$ is a dynamical quantity depending on several factors: time, spatial location, and X-ray luminosity.  
Figure~\ref{fig-alpha_B-Parameter} shows the evolution of $\chi$ in Run 26 as a function of radius and time.  
The spatial distribution (in the left panel) indicates that  
$\chi$ is non-zero only within certain $r$-range, where there is cold gas.  
In this run with $L_X = 0.01$, the cold gas starts appearing within $r \sim 5 - 20$ pc, where cooling instability occurs.  
The range increases to $r \sim 1 - 30$ pc at later times.  
The Bondi radius of the hot gas (with $T = 10^6$ K) is $\sim 18$ pc.  
There is no cold gas within $r < 1$\,pc due to the strong adiabatic heating near the center. 
The maximum value of $\chi$ varies between $1 - 5$.  

We also examined the temporal evolution of $\chi$ at $r=10$\,pc by counting all particles whose smoothing kernel touches $r=10$\,pc.  
The result is plotted in the right panel of Figure~\ref{fig-alpha_B-Parameter}, showing that $\chi$ increases with time.  
The value of $\chi$ increases quickly during the first 0.2 Myr, and after that it remains almost constant at a value $\chi \sim 3 - 4$.


\subsection{$L_X = 0.02$: Cold Clumps and Hot Bubbles}           
\label{sec-LX-0.02}   

Run 27 shows more complex non-spherical behavior in the form of cold clumps and buoyantly rising hot bubbles.   
The fragmentation and clumping start to develop from $t\sim 1.7$\,Myr, and continues until the simulation ends at $t=2.46$\,Myr.   
The gas cools and becomes denser over $r \sim 40 - 100$\,pc,   
following the physics of TI described in \S\ref{sec-TI}.      
This radial range where TI-induced fragmentation occurs is at larger radii than in Run 26 $(L_X = 0.01)$.    
Because with twice higher $L_X$ in Run 27, the $\xi$-range for TI      
(steepest part of the $\Trad - \xi$ equilibrium curve) is satisfied at larger radii.    

Figure~\ref{fig-Splash-Lx0.02-100pcXZ} shows the time evolution of inner $100$\,pc cross-section   
of Run 27 in the $[x - z]$ plane through $y = 0$.      
The gas is heated up and expands into a spherically-symmetric outflow in the inner volume ($r < 40$ pc),   
as the top row at $t = 1.86$\,Myr illustrates.  
This initial outflow decreases the mass inflow rate at $r_{\rm in}$   
by a factor of $\sim 1000$ at $t=1.4 - 2.2$\,Myr (discussed in \S\ref{sec-AccretionModes}).
The surrounding gas at $r=40 - 70$\,pc is inflowing, and has fragmented in certain regions.       
The central outflow collides with this clumpy inflow at $r \sim 40$\,pc, creating a shock.   
It is visible as a ring-like density enhancement, which corresponds to the pile-up of gas   
outflown from the center and the cold clumps from outer regions trying to move in.   

The dense clumps forming at $r \geq 40$\,pc    
continue to move inward, disrupting the spherical-symmetry of the central hotter outflow.    
By $t = 2.12$\,Myr (middle row of Figure~\ref{fig-Splash-Lx0.02-100pcXZ}),     
the whole volume of $r < 80$\,pc shows non-spherical features.   
We see the formation of a hot, less-dense bubble moving outward along the positive $x$-axis just off-center.      

The clumps finally start accreting into $r_{\rm in}$ at $t \geq 2.2$\,Myr.    
This causes the central mass inflow rate in this run to rise again. 
The bottom row ($t = 2.46$\,Myr) of Figure~\ref{fig-Splash-Lx0.02-100pcXZ}   
shows an inhomogeneous multi-phase gas motion.    
There are signatures of a few hot bubbles buoyantly rising,    
which are visible as the density depressions in the positive-$x$ plane.     
The bubbles are not well developed here as compared to Run 28 (\S\ref{sec-LX-0.05}),      
because during their propagation they encounter clumpy medium and become distorted.


\subsection{$L_X = 0.05$: Hot Bubble}                      
\label{sec-LX-0.05}  

The gas is heated up strongly by the highest $L_X$ in Run 28, and there is a spherical outflow.  
There is as well as an anisotropic hot buoyant bubble    
developing because of convective instability (timescales compared in \S\ref{sec-Convection-Time}).   
A large fraction of gas is ejected out of the computational volume,  
and the central mass inflow rate drops drastically (\S\ref{sec-AccretionModes}).  
Convective instability becomes important in accretion flows locally dominated by external heating,   
primarily in regions of subsonic flow \citep{Balbus89}.   
Therefore this dynamical instability is prominent in our larger $L_X$ runs,   
where the gas is heated up and subsonic.    

A spherically-symmetric outflow develops inside of $r \sim 20$\,pc from $t = 1.46$\,Myr,  
when the outer volume still has a spherical inflow.  
At $t = 1.75$\,Myr there is still a $30$\,pc wide shell of inflowing gas at $r \sim 140$\,pc.  
The gas is outflowing all over the volume after $t = 1.8$\,Myr.  

Figure~\ref{fig-Splash-Lx0.05-200pcYZ} shows the cross-section of the whole computational volume  
$\pm 200$\,pc of Run 28 in the $[y - z]$ plane through $x = 0$ at two time epochs.  
There is the formation of a well-defined hot, less-dense bubble,  
which buoyantly rises from the center along the negative $z$-axis.  
In Figure~\ref{fig-Splash-Lx0.05-200pcYZ} it corresponds to the anisotropic jet-like structure:  
redder in temperature-plane, and  bluer in density-plane.  
It starts at $t \sim 1.7$\,Myr as a temperature enhancement and density depression at $z \sim -1$\,pc.  
The top row at $t = 1.8$\,Myr shows the bubble near its formation.  
It grows with time, with gas moving faster inside it than the surroundings.  
The bottom row at $t = 3.0$\,Myr shows a developed stage of the bubble.  
It has a jet-like structure, being fed by hot gas near $r_{\rm in}$.  
Its head presents a bow shock-like morphology on top of a mushroom cloud;  
the interior gas collides with the relatively cold, ambient medium and produces a shock larger than the jet.  
The bubble head reaches  $r_{\rm out}$ at $t \sim 3.8$\,Myr, and the rest of it dissipates after that.  

In this run the remaining of the gas outflow is spherically-symmetric.
Propagating through relatively uniform ambient gas, the bubble is hence well developed here,  
as compared to the distorted ones in Run 27.

The properties of gas from two characteristic regions of the computational volume in this run  
at $t = 2.0$\,Myr is plotted in Figure~\ref{fig-Scatter-Lx0.05}.  
The cyan plus symbols are particles at $r < 60$\,pc overlapping with the negative-$y$ axis  
through their smoothing lengths, hence correspond to part of the bubble.  
The black plus symbols are particles overlapping with the positive-$z$ axis,   
representing the gas in the remaining volume, and show spherically-symmetric behavior.  

The black points can be compared to Figure~16 in Paper~I, which shows the same simulation run   
at an earlier time ($t = 1.5$ Myr) where the flow is still spherically-symmetric,  
and the spherical outflow had already developed.  
The inner gas in Figure~16 is pushed outward continuously,  
and the points are absent at $r < 1$\,pc here in Figure~\ref{fig-Scatter-Lx0.05}.  

Outside the bubble the radial velocity shows mostly outflowing gas, with some particles inflowing at $r < 10$\,pc.  
The strong outflow in this run produces a density profile declining with decreasing radius,  
which is an opposite radial slope compared to all other runs (e.g., Figure~\ref{fig-Scatter-Lx0.01} top-right panel).  
The temperature and entropy plots show increasing heating at smaller $r$, as expected by the high $L_X$.  
In the temperature versus ionization parameter planes ($T - \xi$ and $T - \Xi$),  
the particles are near the hot, stable branch of the radiative equilibrium curve.  
But they are below $\Trad$ since the outflowing gas is adiabatically cooling.  

The bubble (cyan points) has greater outward velocity, lower density, higher temperature and higher entropy  
compared to the gas at the same radius. 
The gas inside the bubble has roughly constant density.  
It also has a roughly constant entropy,  
indicating that indeed the bubble is buoyantly rising and convectively unstable.  
There is a slight decrease in temperature with increasing radius, which is an indication of adiabatic cooling.  

The time evolution of a single particle inside the bubble  
is shown as the blue curve overplotted in each panel of Figure~\ref{fig-Scatter-Lx0.05},  
accompanied by the letters A$-$G in blue.  
The starting point A is indicated by the triangle, which is from the IC at $t = 1.4$\,Myr and $r = 11$\,pc.  
The ending point G denoted by the square is at $t = 3.8$\,Myr when it is at $r = 163$\,pc,  
and the direction of evolution is shown by the arrow.  
The particle is heated up from A to E at varying rates before reaching the bubble at point E, but finally cools inside it from E to G.  
Apart from an initial rise from A to C, the density decreases monotonically from C to G at different rates.  
The entropy increases at first from A to E, and becomes almost constant inside the bubble from E to G.  

There is an initial quick radiative heating from A to B, since the IC has been taken from a run with a lower $\xi$, after which the particle reaches an unstable branch of $\Trad$ at point B.  
It then undergoes heating instability from point B, because of the high-$L_X$ in this run,  
and tries to reach the upper hot-stable branch.  
Once it becomes a part of the bubble at point E, its temperature remains almost constant as it is pushed out fast from E to F.  
It cools adiabatically while moving outward from F to G.  
The position of the particle at $t = 2.0$ Myr is denoted by the orange filled circle.

\subsubsection{Timescale Comparison}   
\label{sec-Convection-Time}    

The outflowing gas is always subsonic in this run, hence is prone to TI if the timescales are appropriate.   
We estimate the outflow motion timescale as: $t_{\rm flow} =  r / | v_r |$,   
and compare it with the TI growth time ($t_{gr}$, Eq.~\ref{eq-tGrowth}).   
The flow is stable initially (we find negative growth rate $n_1$) at simulation times $\leq 1.6$ Myr,    
and hence the gas do not show prominent clumping like in the previous runs.   
The flow becomes unstable later, as we find $t_{gr} < t_{\rm flow}$ within a range $r \sim 100 - 120$ pc   
occurring between a time interval $\sim 1.6 - 2$ Myr.   
We argue that TI comes to play at that epoch leading to the formation of a cold,   
dense spherical shell (which is clumpy at the beginning),    
which forms at $\sim 100$ pc and moves outward.  
The shell can be seen in the top right panel of Figure~\ref{fig-Splash-Lx0.05-200pcYZ}, and clearly in the    
movie\footnote{http://adlibitum.oats.inaf.it/barai/AllPages/Images-Movies/2011/SphericalAccretion/gamma\_5by3\_Run37/LxByLedd\_5e-2/Tgas.mov}.  

We verify the convective origin of the bubble by comparing the relevant timescales.  
We compute the Brunt-Vaisala timescale $(t_{BV})$ for convective instability   
using the effective Brunt-Vaisala frequency $(\omega_{BV})$ from Equation~(4.4) of \citet{Balbus89},   
\begin{equation}  
\label{eq-tBV}   
t_{BV} = \frac{1} {\omega_{BV}} = \left[ \left( - \frac{1}{\rho} \frac{\partial P}{\partial r} \right)   
\left( \frac{3}{5} \frac{\partial \ln P}{\partial r} -  \frac{\partial \ln \rho}{\partial r} \right)  \right]^{-1/2}.   
\end{equation}   
It is estimated when the flow is nearly steady     
and spherical, using the median values of gas density and pressure at each $r$.   
The ratio of $t_{BV}$ with the free-fall time is plotted in Figure~\ref{fig-BruntVaisalaTime-FF}.   
We find $\omega_{BV}^2 > 0$ in some radial ranges indicating convectively unstable gas.   
At the other radii we see convectively stable gas, where $\omega_{BV}^2 < 0$,   
indicated by the radial gaps in Figure~\ref{fig-BruntVaisalaTime-FF}.   

Within the unstable zone, we see that $t_{BV}$ becomes smaller than $t_{ff}$ at some $r$   
which can be attributed to the development of the outflow and bubble in this run.   
In Figure~\ref{fig-BruntVaisalaTime-FF}    
we find a convective instability point at $r \sim 15$ pc at simulation time $1.46$ Myr where $t_{BV} < t_{ff}$,   
which propagates outward with time.  
This represents the spherically symmetric outflow.   
There are also certain radii in the range $(0.1 - 2)$ pc where $t_{BV} < t_{ff}$,  
in between simulation times $1.5 - 1.7$ Myr.   
One such occurrence of convective instability is responsible    
for the development of the anisotropic bubble.


\subsection{Comparing All Runs} 
\label{sec-AccretionModes} 

We observe various modes of gas accretion for different values of $L_X$. 
The global features of the flow in each run are summarized in Table~\ref{Table-AccrModes}.  
In Run 25 with $L_{X} = 0.005$, there is a stable inflow maintaining spherical-symmetry. 
This run is listed in the table for comparison, and has been discussed in Paper~I.    
The gas heats up as $L_X$ is increased,  
and the accretion flow undergoes unstable motion causing non-spherical features. 
With $L_X = 0.01$ and $0.02$ 
the gas cools and fragments forming multiple cold, dense clumps and attains a multi-phase structure. 
In the runs where a spherically-symmetric outflow is produced in the inner volume ($L_X = 0.02, 0.05$) initially, 
there is also the formation of hot bubbles rising buoyantly from the center.    

Figure~\ref{fig-Mdot_rin} shows the time evolution of the     
mass inflow rate at the inner radius ($\dot{M}_{{\rm in}, r_{\rm in}}$) for the four runs.  
We see that $\dot{M}_{{\rm in},r_{\rm in}}$ drops at a varying rate as $L_{X}$ is increased,   
because the gas is heated up and expands.   
The inflow continues in the inner volume ($r < 40$\,pc) with $L_X = 0.005$ and $0.01$ at a rate $2 - 4 \, M_{\odot}$/yr,   
with $\dot{M}_{{\rm in}, r_{\rm in}}$ reduced by a factor of $0.84$ in the latter run.  
When $L_X = 0.02$ or higher, a spherical outflow develops at $r < 40$\,pc,   
because the gas is efficiently heated up by the central radiation.        
This outflow suppresses the accretion rate by $3 - 4$ orders of magnitude between $1.4 - 1.6$\,Myr in Runs 27 and 28.  
The mass inflow rate in Run 27 rises again after $2.2$\,Myr,  
because of the accretion of cooler clumps (details in \S\ref{sec-LX-0.02}).  
There is however a strong outflow with $L_{X} = 0.05$ which expels a substantial fraction of gas  
out of the computational volume, and $\dot{M}_{{\rm in}, r_{\rm in}}$ never rises within $5$\,Myr.  


With $L_X = 0.01$ (Run 26) the hotter outflow reaches   
a radial velocity $v_{\rm out} \leq 300$\,km/s between $10 - 30$\,pc.   
In Run 27 ($L_X = 0.02$), initially there is a spherically-symmetric outflow with $v_{\rm out} \leq 200$\,km/s at $r < 40$\,pc.   
The outflow becomes stronger at $t \geq 2.0$\,Myr when hotter gas and buoyant bubbles become channelled   
in-between the inflowing cool, dense clumps, reaching $v_{\rm out} \leq 800$\,km/s.   
With $L_X = 0.05$ in Run 28, the spherical outflow has $v_{\rm out} \leq 200$\,km/s, and   
the anisotropic hot buoyant bubble consist of faster moving gas with $v_{\rm out} \leq 600$\,km/s.  
In runs with $L_X = 0.02$ and $0.05$ the fastest outflowing gas is subsonic at the beginning,   
and becomes supersonic later with Mach number of $2 - 3$.   
At $t \geq 1.8$\,Myr there is always some gas at $v_{\rm out} \sim 100 - 200$\,km/s which is supersonic.  
Initially most of the gas has $v_{\rm out}$ smaller than the escape velocity ($v_{\rm esc}$). 
At $t > 2$\,Myr some gas between $40 - 120$ pc flows out with $v \ge v_{\rm esc}$.   

The mass outflow rate $(\dot{M}_{\rm out})$ as a function of radius at a few specified times,   
and its time evolution at a given radius for the three runs are plotted in Figure~\ref{fig-Mdot_Outflow}.   
The outflow rate is low with $L_X = 0.01$, only reaching $\dot{M}_{\rm out} \leq 0.1 M_{\odot}$/yr at $r \sim 20$\,pc.   
It increases in the higher $L_{X}$ runs, because of the development of strong outflows.   
With $L_X = 0.02$ it rises up to $\dot{M}_{\rm out} \sim 1 M_{\odot}$/yr between $r = 20 - 80$\,pc.   
For $L_X = 0.05$ it reaches $\dot{M}_{\rm out} = 5 - 10 ~ M_{\odot}$/yr at $r = 40 - 100$\,pc.   
The results in our GADGET runs with $L_X = 0.01$ and $0.02$   
are roughly in the same range as that obtained in the ZEUS counterpart simulations, which gave 
$\dot{M}_{\rm out} \sim 0.1 - 1 ~ M_{\odot}$/yr \citep{KP09a}.   

\section{Discussions}  
\label{sec-discussion}  

Our analysis of the hot versus cold gas fractions in the accretion flow (\S\ref{sec-Cold-Hot-Frac})  
can be used to constrain the $\alpha$-parameter in the expression for SMBH accretion rate  
used in the sub-grid models of cosmological simulations.  
In such models, the Bondi-Hoyle-Lyttleton mass accretion rate \citep{Hoyle39, Bondi44, Bondi52}  
inferred from simulations is multiplied by the factor $\alpha$, with a value of the order $\alpha = 100$  
\citep[e.g.,][]{DiMatteo08, Dubois10, Lusso10}.  
One of the reasons given for such high values of $\alpha$ is the artificially low densities  
and high temperatures of the ISM obtained by smoothing on the scale of 
computational resolution at the location of the BH \citep{Booth09, Johansson09, Khalatyan08}.  
Another reason given is that a volume-average of the Bondi-rates for the (spatially unresolved)  
cold and hot phases of the ISM recovers a value of $\alpha$ close to $100$ \citep{Sijacki09}.  
Effectively the galaxy simulations need to assume that a large fraction of unresolved cold gas
accretes onto the central SMBH, in order to grow the BHs to currently observed masses. 

However we do not find a factor as large as $100$ for the cold gas fraction, as our $\chi \leq 1 - 5$.  
In our simulations we do not see a large amount of gas hidden in the cold phase.  
Other physical processes, like self-gravity and rotation which we do not have in our current simulations,  
might contribute to the conversion of hot accreting gas to the cold phase and increase $\chi$. 
The value of $\alpha$ should be a resolution dependent quantity.  
A self-consistent determination of $\alpha$ needs multi-leveled high-resolution approach  
to study both star formation and gas accretion onto SMBH (see recent attempt by \citealt{Kim11}).   

The mass inflow rate (Figure~\ref{fig-Mdot_rin}, \S\ref{sec-AccretionModes})    
shows large fluctuations in Run 26 ($L_X = 0.01$) after $1.6$\,Myr, and in Run 27 ($L_X = 0.02$) after $2.3$\,Myr.   
This is because of the accretion of multiphase gas,  
with the spikes corresponding to the infall of cold dense clumps.  
It is similar to the result by \citet{Sharma11}, who found large variations of the mass accretion rate  
as a function of time when the cold TI-induced filaments cross the inner boundary.  
The clumps in our simulations have density between $10^{-22} - 10^{-20}$\,g/cm$^3$, 
which overlap with the upper limit where NLR clouds are observed having    
electron density between $10^2 - 10^6$\,cm$^{-3}$ (mass density $10^{-25} - 10^{-21}$\,g/cm$^3$).   
The temperature of the simulation clumps is between $10^4 - 10^5$\,K at $r \ge 1$\,pc,    
whose lower limit coincides with the observed NLR temperature range of $10^4 - 2.5 \times 10^4$\,K.   
The outward velocities obtained in our simulations (few $100$ km/s) are smaller than the   
upper limit of few $1000$\,km/s observed in X-ray spectroscopic studies of AGN warm absorber outflows   
\citep{Turner05, Evans07, Markowitz08, Turner08, Mehdipour10, Turner10},   
which could be due to the absence of radiation pressure in our current simulations. 

Thermal conduction becomes important in order to describe the physical evolution of TI  
and to resolve the critical length scale of TI \citep[e.g.,][]{Koyama04}.  
The relevant scale is the Field length ($\lambda_F$), below which thermal conduction  
suppresses the growth of density perturbations due to TI by efficiently conducting heat.  
We compute $\lambda_F$ for our cooling function,  
finding that it increases monotonically with increasing radius and $L_{X}$.  
The highest values are in the run with $L_{X} = 0.05$, which lies between  
$\lambda_F \sim 4 \times 10^{-6}$\,pc near $r_{\rm in}$, and $\lambda_F \sim 0.1$\,pc near $r_{\rm out}$.  
The smoothing lengths in our simulations are 
$\sim 0.07$\,pc near $r_{\rm in}$, and $\sim 10$\,pc near $r_{\rm out}$.  
Thus our spatial resolution limit is at least $10^2 - 10^4$ times larger than $\lambda_F$.  
Since we do not resolve the scales    
where the growth of TI-induced density perturbations is suppressed by heat conduction,   
our simulations can only give an upper limit on the size of the smallest clumps.


In order to check if our simulations can follow the isobaric mode of TI (i.e., perturbations with a constant-pressure),  
we compare the sound crossing time ($t_S$) with the cooling time ($t_C$),   
both local values calculated for each gas particle.  
The GADGET code determines the adaptive smoothing length ($h_{sml}$) of each particle to include  
a constant mass for the estimated density, amounting to $32 - 64$ smoothing neighbors within $h_{sml}$.  
In our simulations we used $50 \pm 1$ as the desired number of SPH smoothing neighbors.  
Therefore within $3 h_{sml}$ there are $\sim$150 neighboring particles,  
which is a typical number needed to resolve a sound wave.  
We determine $t_S$ by computing   
how long it would take a sound wave at a given radius to cross $3 h_{sml}$.   
We find that $t_S > t_C$ everywhere before the gas undergoes TI-driven cooling, with $[t_S / t_C]_{\rm min} \sim 2$.   
\citet{Stellingwerf82} studied the local and global stability of spherically symmetric,   
X-ray heated and optically thin accretion flow.   
He gave a critical wavelength $\lambda_0 = 4 \pi \gamma^{1/2} c_s u / (du/dt)_{\rm Cool}$,  
below which the instability growth time is independent of $\lambda$.   
Here $c_s$ is the sound speed, $u$ is the specific internal energy,   
and $(du/dt)_{\rm Cool}$ is the cooling term in the specific internal energy equation.   
Wavelengths shorter than $\lambda_0$ are most unstable, and such perturbations are approximately isobaric.   
Comparing $\lambda_0$ with $h_{sml}$ in our simulations,      
we see that $h_{sml} < \lambda_0$ before TI-induced cooling starts to happen.   
Hence we can resolve isobaric modes of TI.   

We check the Jeans instability criterion in order to determine  
if the cool, dense clumps formed in our simulations would undergo self-gravitating collapse.  
Our results indicate that self-gravity is sub-dominant over gas pressure.  
Taking the density and temperature of two typical clumps,  
we calculate the corresponding Jeans length ($\lambda_J$) and mass ($M_J$).  
For a filament-like clump with $\rho = 10^{-20}$\,g/cm$^3$, $T = 3.2 \times 10^5$\,K, and of length $\sim 30$\,pc,  
from Figure~\ref{fig-Splash-Lx0.01-30pcYZ}:  
$\lambda_J = 107.9$\,pc, and $M_J = 9.7 \times 10^7 M_{\odot}$.  
For a more-spherical clump with $\rho = 5.0 \times 10^{-20}$\,g/cm$^3$, $T = 3.2 \times 10^6$\,K,  
and of length $\sim 1$\,pc, from Figure~\ref{fig-Splash-Lx0.01-4pcXY}:  
$\lambda_J = 152.4$\,pc, and $M_J = 1.4 \times 10^9 M_{\odot}$.  
As a comparison the initial mass in the whole computational volume is  
$M_{\rm tot,IC} = 9.77 \times 10^{6} M_{\odot}$ (Table~\ref{Table-AccrModes}).  
Therefore the computed Jeans length and mass are greater than the actual length and mass of the clumps, 
which means they would not self-gravitate, and gas pressure would restore balance over self-gravity.   

We compare our results to those from the Eulerian code ZEUS, where spherical-symmetry can be maintained to a higher degree.  
Results from 1D and 2D ZEUS counterpart simulations show the formation of  
cold, dense, spherically-symmetric shells due to TI.     
These shells form continuously, at non-uniform time intervals,   
accrete in and cause large spikes in the central mass accretion rate.  
However the flow remains spherical in ZEUS.   
We deduce that small non-spherical density perturbations present in SPH (from the initial condition),   
which are random, grow via TI resulting in the inhomogeneous flow presented in \S\ref{sec-results}.  
This has been tested by adding non-spherical density fluctuations to the initial condition of ZEUS simulations;  
then the flow becomes clumpy and inhomogeneous in ZEUS as well.   
Therefore we cannot study the clumps in a very controlled manner using SPH.   

Examining the parameter space of the results described here, we see the occurrence of TI-induced clumping  
for a relatively narrow range of X-ray luminosity ($L_X = 0.01 - 0.05$).  
But this is only for a fixed density ($\rho_{\infty} = 10^{-23}$ g/cm$^3$).  
At luminosities below $L_X = 0.01$ stable spherically-symmetric gas inflow occurs,   
and above $L_X = 0.05$ there is a spherical outflow. 
The range of photoionization parameter ($\xi = \frac{L_X}{r^2 n}$)   
for the operation of TI depends on $L_X$, $r$ and $n$.  
We also observe non-spherical cooling in a run with $\rho_{\infty} = 10^{-24}$\,g/cm$^3$, and $L_X = 5 \times 10^{-4}$.  
This supports our assertion that such fragmentation corresponds to the accreting gas   
attaining the $\xi$-range for TI (largest slope of $[\Trad - \xi]$ curve, or $dT / d\Xi < 0$, \S\ref{sec-TI}).  

\citet{Krolik88} invoked a finite-amplitude TI for the origin of broad emission line clouds in AGN.  
His Figure~1 shows the operation of TI in the $[T - \Xi]$ plane.  
However in our simulations the initial conditions of the gas are different, therefore we see a different time evolution.   
We rather find a typical $T - \Xi$ transition of the kind shown in the   
bottom-right panel of Figure~\ref{fig-Evolution-Single} and described in \S\ref{sec-TI}, for the particles that experience TI.  


Our results show that fragmentation can occur in a simple system with minimal physical processes:  
a spherically accreting gas can cool non-spherically and become clumpy due to thermal instability 
and radiative heating/cooling. 
There is of course the additional contribution of small density fluctuations inherent to SPH.  
At the same time, our analyses using the mesh code ZEUS         
indicate that when a steady solution is very unstable, it is hard to avoid the development of TI.   
The main difference with various numerical techniques is the amplitude and type of perturbation.  
Operating in such a manner TI can induce, or at least enhance,   
star-formation near AGN, by causing the gas to cool and fragment.  
Global TI can be a triggering mechanism for the variability observed in the luminosity of compact objects,  
by making the accreting gas inhomogeneous.   

In our simulations the gas undergoes spherical accretion initially, and  
the radiation from the central SMBH is modeled in the form of a spherical X-ray corona. 
Possible future work would include breaking such spherical symmetry,  
e.g., considering radiation from an optically thick, geometrically thin, standard accretion disk \citep{Proga07},  
including rotation of the accreting gas \citep{POK08}, and radiation force from the disk \citep{KP08}.


\section{Summary and Conclusion} 
\label{sec-conclusion}   

Gas accretion onto a central SMBH was studied in Paper~I  
by performing simulations of spherical Bondi accretion using the 3D SPH code GADGET-3.   
The simulations included radiative heating by a central X-ray corona and radiative cooling.  
All the results presented in Paper~I were of spherically-symmetric gas properties.  
However, for a certain parameter range of  
X-ray luminosity $L_{X}/L_{\rm Edd} = 0.01, 0.02, 0.05$ with a fixed density at the outer boundary   
$\rho_{\infty} = 10^{-23}$\,g/cm$^3$ (i.e., within the unstable $\xi$-range, \S\ref{sec-TI}),    
we observe non-spherical, multi-phase gas motion.  
In the current work we present detailed analyses of the inhomogeneous gas behavior. 
The main results are summarized below: 

(1) The gas goes through various modes of accretion for different $L_{X}$.  
There is a stable, spherically-symmetric inflow with $L_{X} = 0.005$ (Run 25).  
The gas is heated up and expands as $L_X$ is increased, which decreases the central mass inflow at a varying rate.  
For some intermediate $L_X$ the accretion flow undergoes unstable motion developing non-spherical features.  
With high enough $L_X$ a strong outflow is produced, ejecting most of the gas outside.

(2) With $L_X = 0.01$ and $0.02$,   
the gas cools and becomes denser in a non-spherical manner, fragments, and forms multiple clumps.  
The clumping occurs over $r < 30$\,pc for $L_{X} = 0.01$, and $r \sim 40 - 100$\,pc for $L_{X} = 0.02$.  
It starts to develop from $\sim 1.6 - 1.7$ Myr, and continues up to the simulation end.   
The clumps are filamentary in structure, supersonic, and have a lower photoionization.  
Hotter, lighter, subsonic gas exist in-between the filaments, and start to flow outward at later times.  
The clumps undergo a dynamic evolution:  
become stretched, fragment, merge with each other, finally accreting into $r_{\rm in}$.  

The gas particle properties versus radius show large scatters (few $100$s to $1000$)  
at the radial ranges where fragmentation occurs, corresponding to the multi-phase structure of the gas.  
The mass inflow rate shows large fluctuations in the run with $L_X = 0.01$ after $1.6$\,Myr,   
and in the run with $L_X = 0.02$ after $2.3$\,Myr,  
because of the accretion of multi-phase gas, with the spikes corresponding to the infall of cold dense clumps.

(3) As they move inward the cooled clumps remain denser, however heat up to above $10^5$\,K.  
The central heating at $r < 1$\,pc is mostly by adiabatic compression reaching $T > 10^7$\,K at $r_{\rm in}$,  
while radiative heating would dominate at a few pc for some particles.  
The photoionization parameter increases while moving to smaller radii, and the clumps remain supersonic.  
Artificial viscosity heating increases in magnitude toward $r_{\rm in}$,  
however it always remains sub-dominant compared to adiabatic heating.  
This is in contrast to Paper~I due to the different conditions here because of the higher $L_{X}$.

(4) The cooling and clumping is caused by the interplay of thermal instability  
in the inflowing gas acted upon by radiative processes, as well as being adiabatically compressed and heated.  
Gas particles in our simulations undergo TI within the unstable $\xi$-range,  
where the $\Trad - \xi$ equilibrium curve has the steepest slope (between $\xi \sim 100 - 500$),  
and most physically relevant processes (those with constant-pressure or free falling)  
can give rise to perturbations with a shallower slope.  

We detect a distinct mode of cooling/heating in the $[T - \xi]$ plane.  
A particle while being dominated by adiabatic process moves up to higher-$T$  
along the unstable part of $\Trad - \xi$.  
This happens because the adiabatic free-fall relation of $T$ versus $\xi$ ($T_{ff, a} \propto \xi_{ff, a}^2$)  
has the same slope as the radiative equilibrium curve in the unstable regions.  
At a later time the particle has a transition to being dominated by radiative process.  
Then classical TI operates, and it undergoes catastrophic cooling or heating (whichever dominates),  
and moves away from the unstable $\Trad - \xi$ part.  
It finally settles on one of the stable branches where $\Trad - \xi$ has a shallower slope.

(5) Typically the amplitude of temperature and density perturbations versus time  
shows that the perturbations have an exponential growth at first,   
and then come to a saturation for the remaining simulation run.  
The minimum temperature has an exponential decrease initially, indicating the growth of cooling instability.   
It remains almost constant after that at the temperature of the lower stable $\Trad - \xi$ branch,   
where $T$ is a weak function of $\xi$.   
In terms of density the maximum value rises indicating the growth of cooled, dense clumps caused by TI.

(6) We have multi-phase medium (cold gas with $2 \times 10^4$\,K and hot gas with $10^6 - 10^7$\,K)  
at the same radius, forming self-consistently in the simulation volume.  
Furthermore we resolve the Bondi radius of the accretion flow.  
This can be used to constrain the $\alpha$-parameter  that is 
multiplied to the Bondi-Hoyle-Lyttleton mass accretion rate in the sub-grid model of SMBH feedback in large-scale galaxy simulations.  

Considering a limiting temperature of $10^5$\,K between cold and hot phases,  
we compute the ratio ($\chi$) of mass inflow rates of cold gas over hot.  
We find that $\chi$ is a dynamical quantity depending on several factors: time, spatial location, and X-ray luminosity.  
The maximum value of $\chi$ varies between $1 - 5$.  
At a fixed radius, $\chi$ increases with time initially,  
and later remains almost-constant at a value $\chi \sim 3 - 4$.  
Thus we do not find a factor as large as $\alpha \sim 100$, since our $\chi \leq 1 - 5$.  
In other words, in our simulations we do not find such a large fraction of gas hidden in the cold phase that is unresolved in cosmological simulations.

(7) With $L_X = 0.02$ (Run 27) we see both fragmented cold clumps and buoyantly rising hot bubbles.  
Initially there is a central spherical outflow at $r < 40$\,pc,  
which decreases the mass inflow rate at $r_{\rm in}$ by $\sim 1000$ times between $1.4 - 2.2$\,Myr.  
The surrounding inflowing gas between $[40 - 70]$\,pc cools and fragments by TI.  
The dense clumps grow with time, and continue to move inward,  
disrupting the spherical symmetry of the central hotter outflow.  

The clumps start accreting into $r_{\rm in}$ at $t \geq 2.2$\,Myr, causing the central mass inflow rate to rise again.  
The gas motion becomes inhomogeneous with cooler, denser clumps falling in to the center 
but heating up as they move in; and hotter, lighter gas moving out.  
There are also few hot bubbles buoyantly rising from the center.

(8) With a high enough value of the X-ray luminosity ($L_X = 0.05$ in Run 28),  
the gas is heated up strongly, creating a spherically symmetric outflow.  
This expels a large fraction of gas out of $r_{\rm out}$, reduces the central mass inflow rate drastically,  
and produces a gas density profile that decreases towards center. 
There is also the formation of an anisotropic, hot bubble buoyantly rising from the center.  
It has a narrow elongated structure, while its head has a bow shock-like morphology,  
where the interior gas collides with the relatively cold ambient medium and produces a shock region larger than the jet.  
The bubble has a greater outward velocity, lower density, higher temperature and higher entropy  
compared to the gas at the same radius.

\section*{Acknowledgments} 
We are grateful to to V. Springel for allowing us to use the Gadget-3 code. 
This work is supported in part by the NSF grant AST-0807491, National Aeronautics and Space Administration 
under Grant/Cooperative Agreement No. NNX08AE57A issued by the Nevada NASA EPSCoR program, 
and the President's Infrastructure Award from UNLV.   
DP also acknowledges the UNLV sabbatical assistance.    
This research is also supported by the NSF through the TeraGrid resources provided by the Texas Advanced Computing Center.
Some numerical simulations and the analyses have been performed on the UNLV Cosmology Cluster.    
PB also acknowledges support from the FP7 ERC Starting Grant ``cosmoIGM''.

%



\begin{onecolumn}

%

\begin{deluxetable}{cccccc}
\tablewidth{0pt}
\tablecaption{Simulations and Flow Features}
\tablehead{
\colhead{Run} & 
\colhead{$L_{X}$} &
\colhead{Spherical \tablenotemark{b}} &
\colhead{Clumping \tablenotemark{c}} & 
\colhead{Bubble \tablenotemark{d}} &  
\colhead{Motion}  
\\ 
\colhead{No. \tablenotemark{a}} & 
\colhead{[$L_{\rm Edd}$]} & & & & 
\colhead{($r < 40$ pc, $1.4 - 2.3$ Myr)} 
}

\startdata


$25$ & $0.005$ & Yes & No & No & Inflow \\

\hline \\


$26$ & $0.01$ & No & Yes & No & Inflow \\

\hline \\

$27$ & $0.02$ & No & Yes & Yes & Outflow \\

\hline \\

$28$ & $0.05$ & No & No & Yes & Outflow \\

\enddata
\label{Table-AccrModes} 

\tablenotetext{a}{  
The run numbers are from Table 2 of Paper~I.  The common parameter values are quoted below.  
These runs are started using the particle states in Run 23 (Paper~I) at $t= 1.4$\,Myr as the initial condition:  
$\rho_{\rm init} = \rho_{\rm Run23}$, $v_{\rm init} = v_{\rm Run23}$, $T_{\rm init} = T_{\rm Run23}$,  
and changing $L_{X}$ in each case.  
All the runs have $\gamma_{\rm run} = 5/3$, $r_{\rm in} = 0.1$\,pc, $r_{\rm out} = 200$\,pc, $N = 1.24 \times 10^{7}$,  
$M_{\rm tot,IC} = 9.77 \times 10^{6} M_{\odot}$, and $M_{\rm part} = 0.791 M_{\odot}$.  
The IC for Run 23 was generated using $\gamma_{\rm init} = 5/3$, $\rho_{\infty} = 10^{-23}$\,g/cm$^3$,  
$T_{\infty} = 10^5$\,K (for which the Bondi radius is $R_B = 183.9$\,pc),  
$\rho_{\rm init} (r) = \rho_B (r)$, $v_{\rm init} (r) = v_B (r)$, and $T_{\rm init} = \Trad$.  
}  
\tablenotetext{b}{ Whether the gas properties in the flow remain spherically-symmetric.} 
\tablenotetext{c}{ Whether the gas undergoes non-uniform cooling, fragmentation and form clumps.} 
\tablenotetext{d}{ Whether there is the formation of hot bubble(s) rising buoyantly from the center.}   

\end{deluxetable}


\clearpage


\begin{figure} 
\centering 
$ 
\begin{array}{cc} 
\includegraphics[width = 0.5 \linewidth]{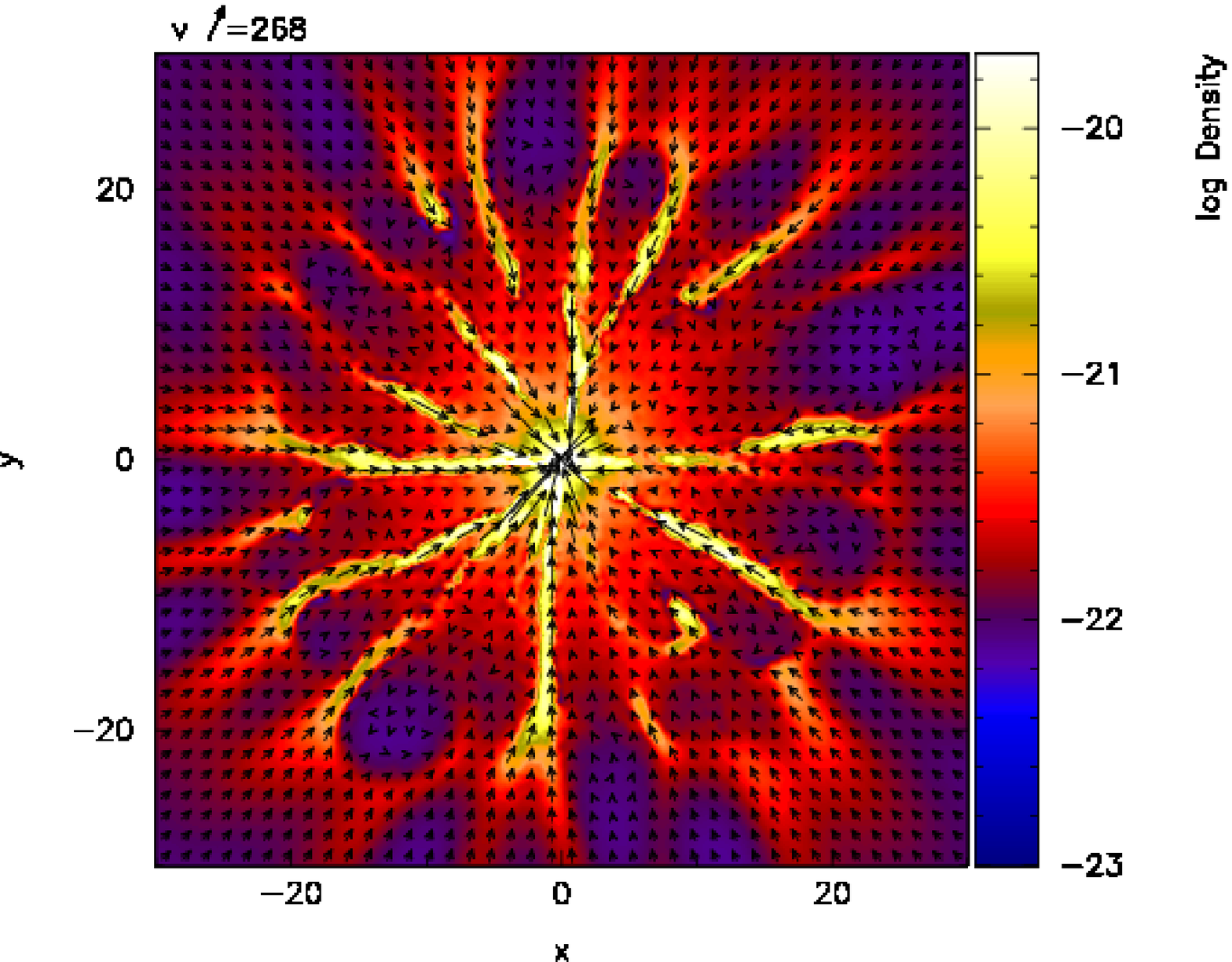} & 
\includegraphics[width = 0.5 \linewidth]{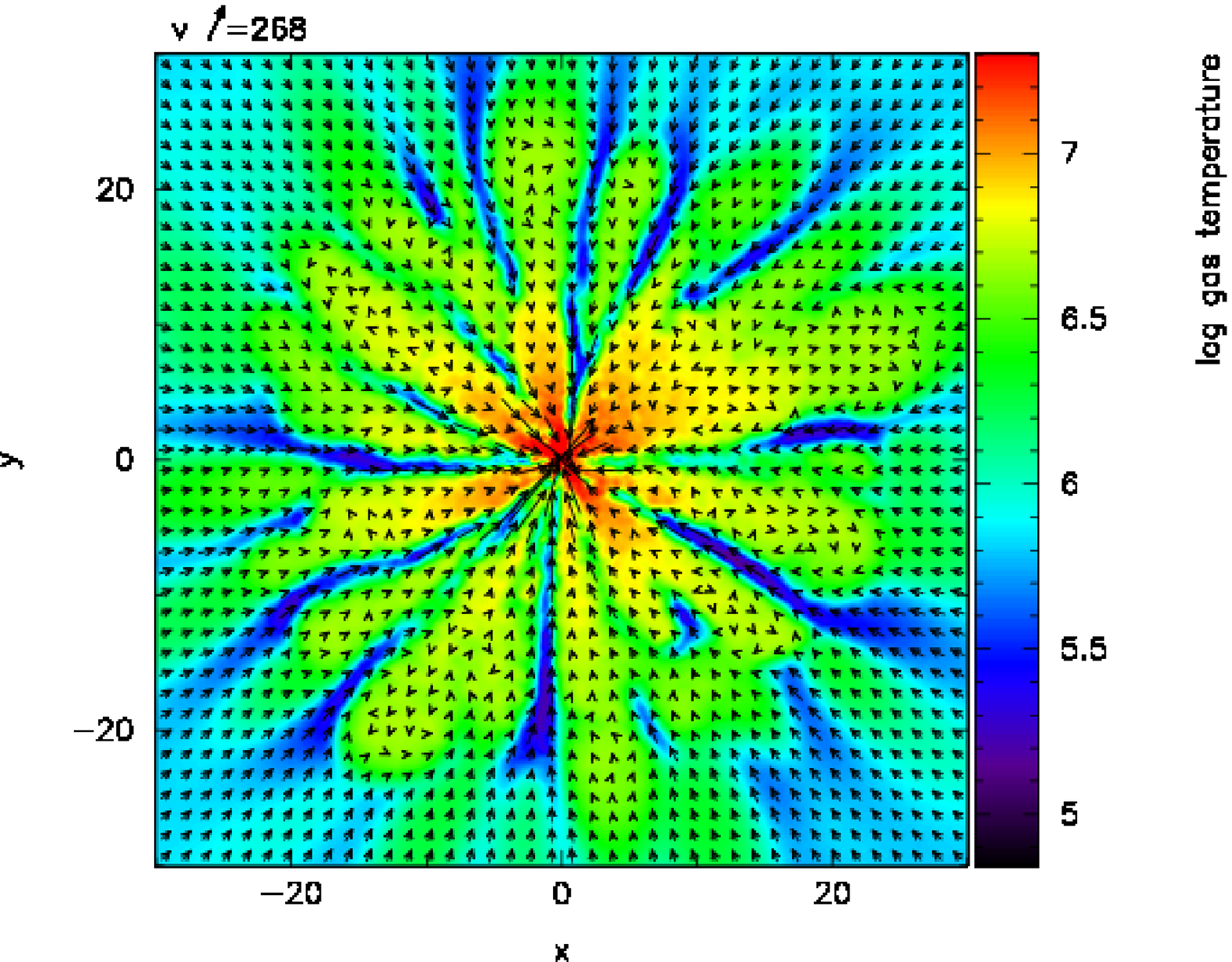} \\  
\includegraphics[width = 0.5 \linewidth]{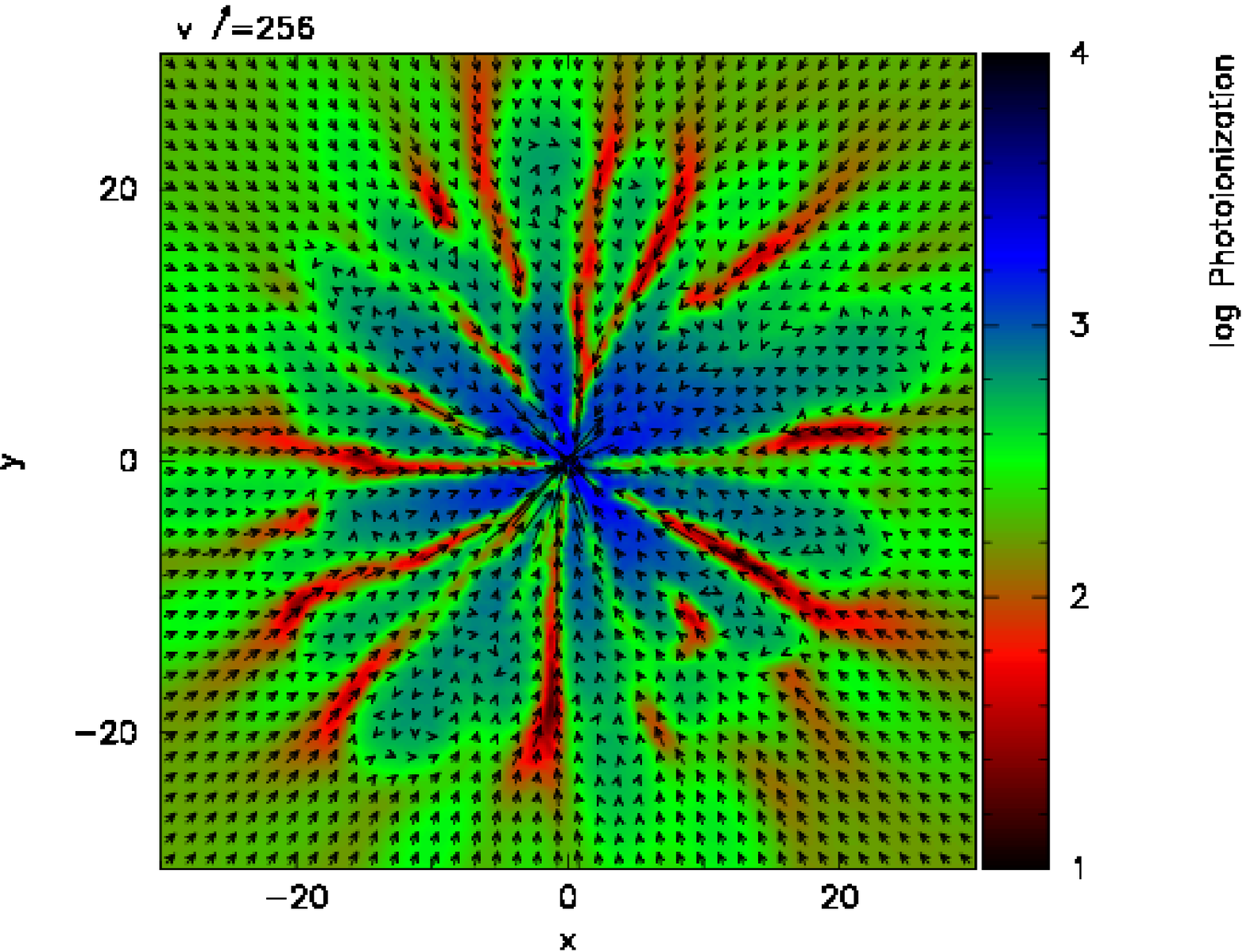} &  
\includegraphics[width = 0.5 \linewidth]{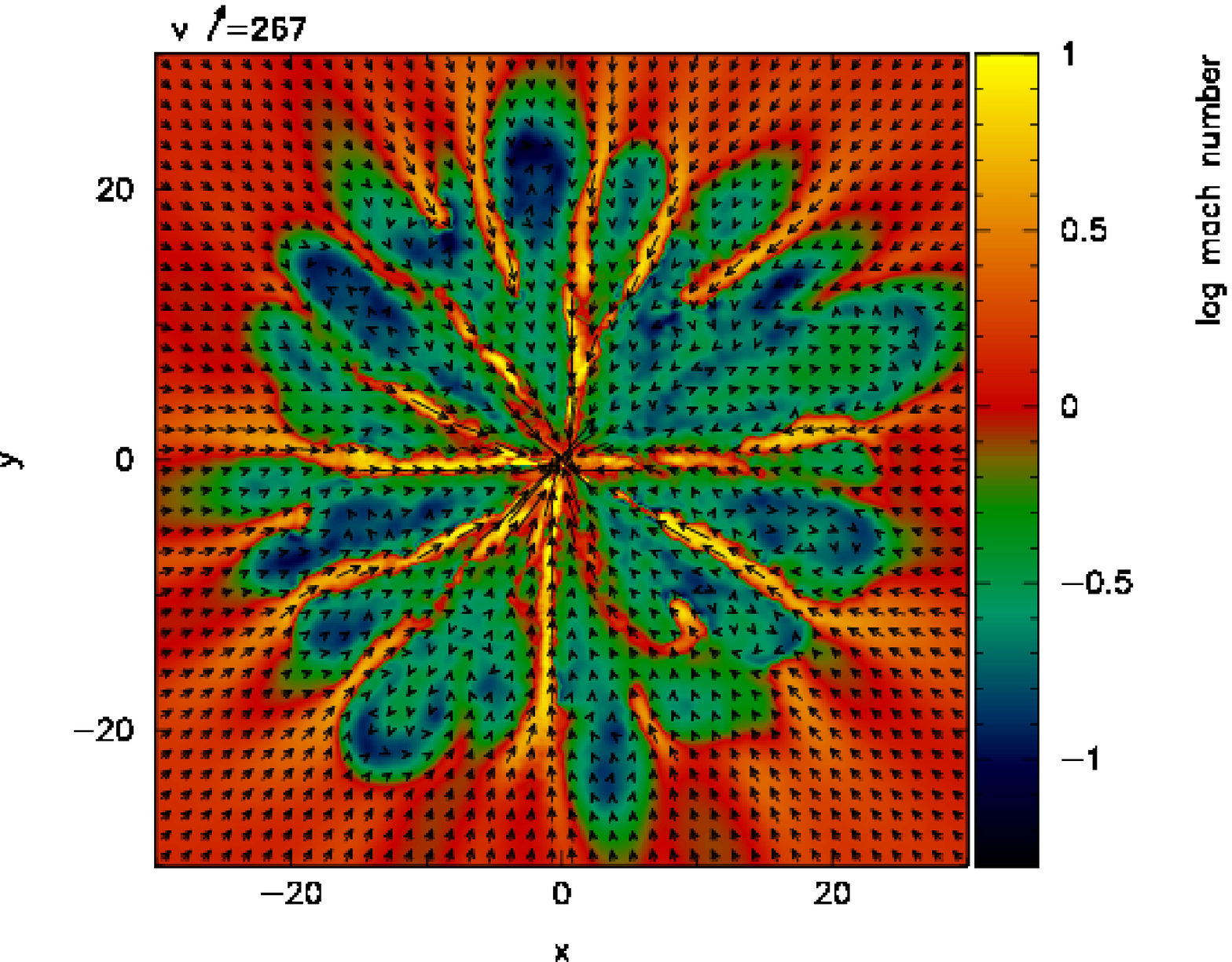}  
\end{array}  
$ 
\caption{ 
Cross-sections in Run $26$ ($L_{X} / L_{\rm Edd} = 0.01$) 
showing the inner $30$\,pc of the $[x - y]$ plane through $z = 0$ at a time $t = 2.047$\,Myr.  
The gas density is in the top-left panel, temperature in the top-right,  
photoionization parameter in the bottom-left, and Mach number in the bottom-right, all in cgs units.   
The velocity vectors are overplotted as arrows,   
whose sizes are denoted at the top-left of each panel in km/s.   
It shows colder, denser filament-like structures, with hotter, less-dense gas in-between,  
both components accreting in (with the colder phase moving in faster).
These features are caused by TI-induced non-spherical cooling and fragmentation.   
Note that a different color scheme has been used to depict each quantity,   
therefore the map between the colors and the order of values (increasing/decreasing)   
could be different from one panel to the other. 
This and all the other cross-section images in this paper have been generated using SPLASH \citep{Price07}.  
}
\label{fig-Splash-Lx0.01-30pcYZ}
\end{figure}

\begin{figure} 
\centering  
\includegraphics[width = 1.03 \linewidth]{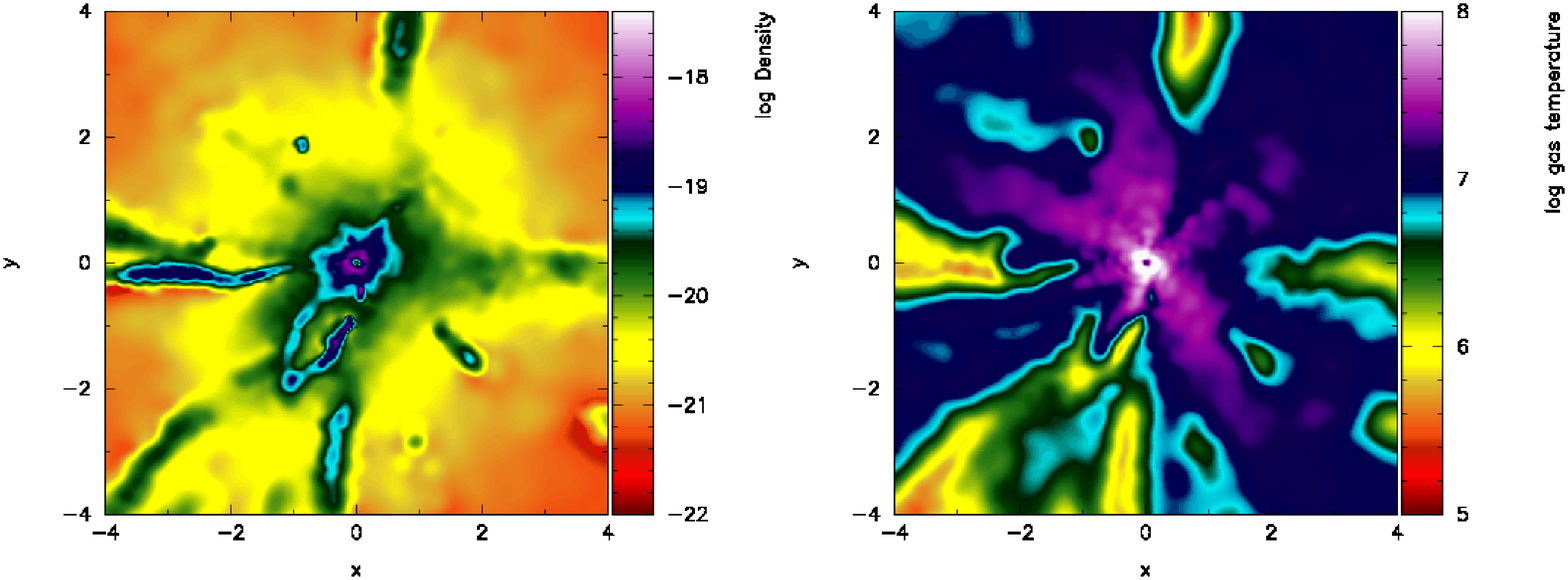} \\  
$ 
\begin{array}{cc} 
\includegraphics[width = 0.5 \linewidth]{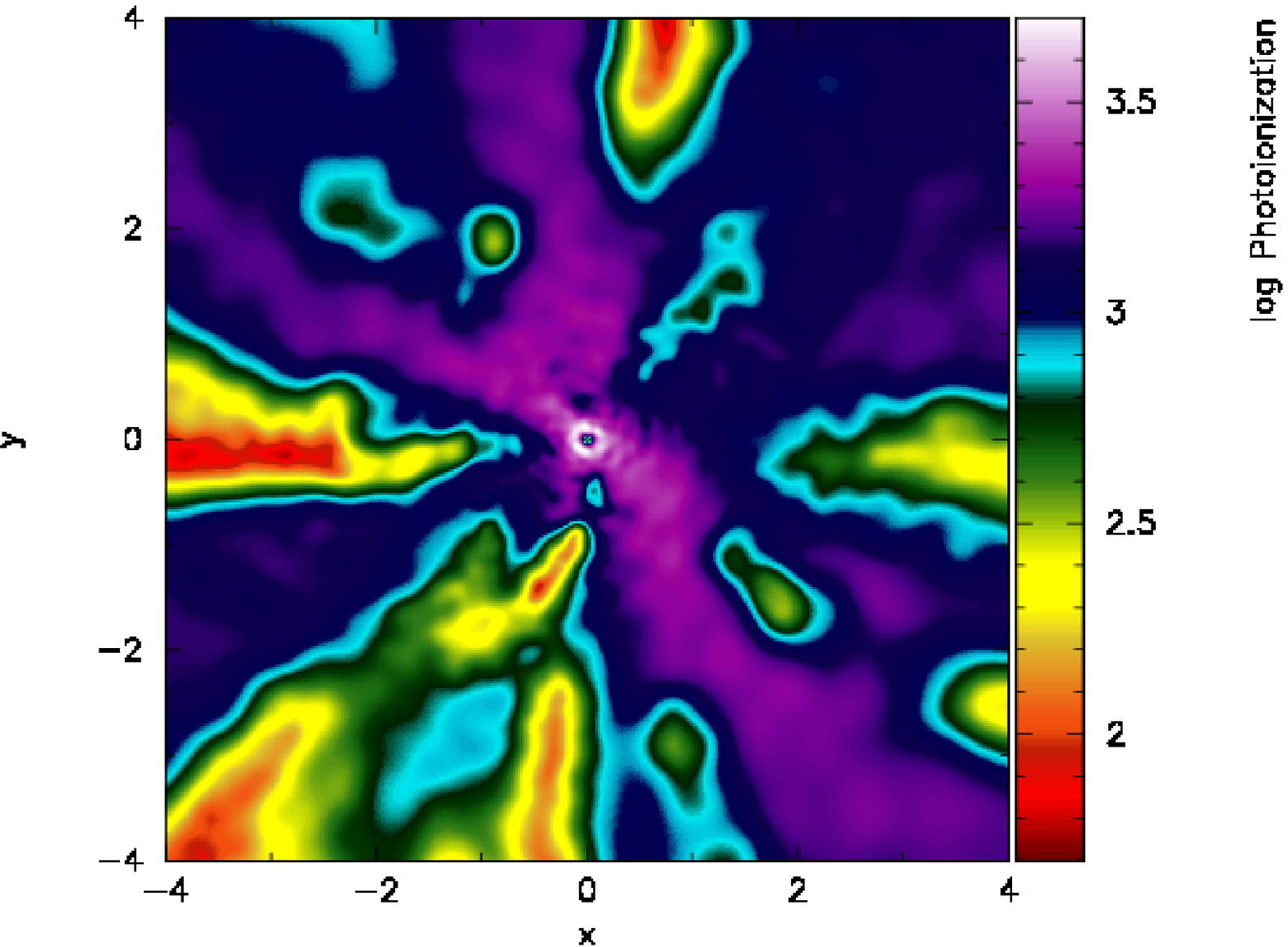} & 
\includegraphics[width = 0.5 \linewidth]{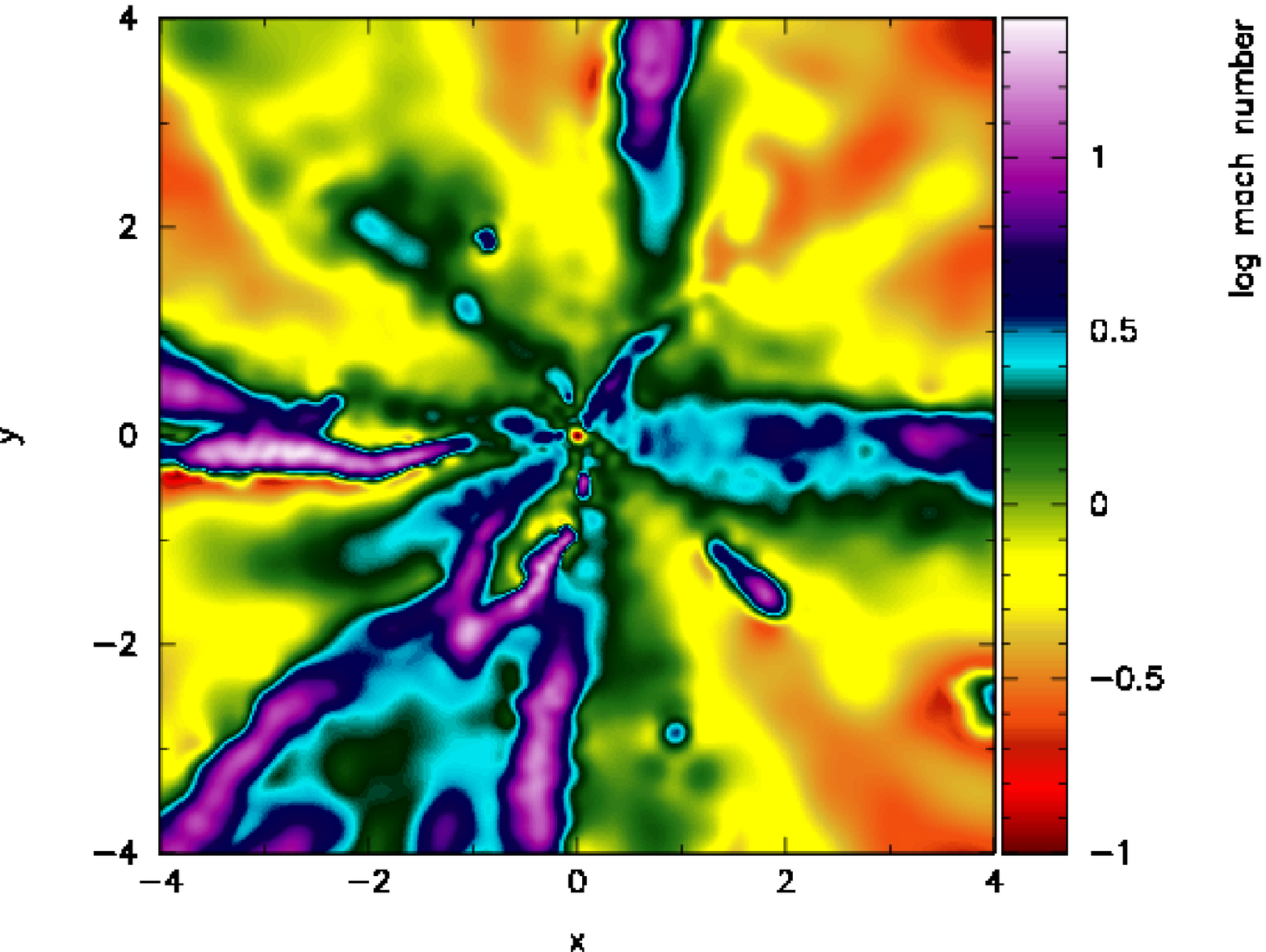} 
\end{array} 
$ 
\caption{  
Same as Figure~\ref{fig-Splash-Lx0.01-30pcYZ} but showing zoom-in of the inner $4$\,pc of the $[x - y]$ plane.  
It shows stretching of the colder clumps as they fall in toward the center.  
They remain denser, however heats up at $r < 1$\,pc, mostly by adiabatic compression.  
Note that the color scheme in this cross-section has been changed and  
it has been plotted without the velocity vectors, in order to show the small-scale features clearly.  
}
\label{fig-Splash-Lx0.01-4pcXY}
\end{figure}

\begin{figure}
\centering
\includegraphics[width = 0.8 \linewidth]{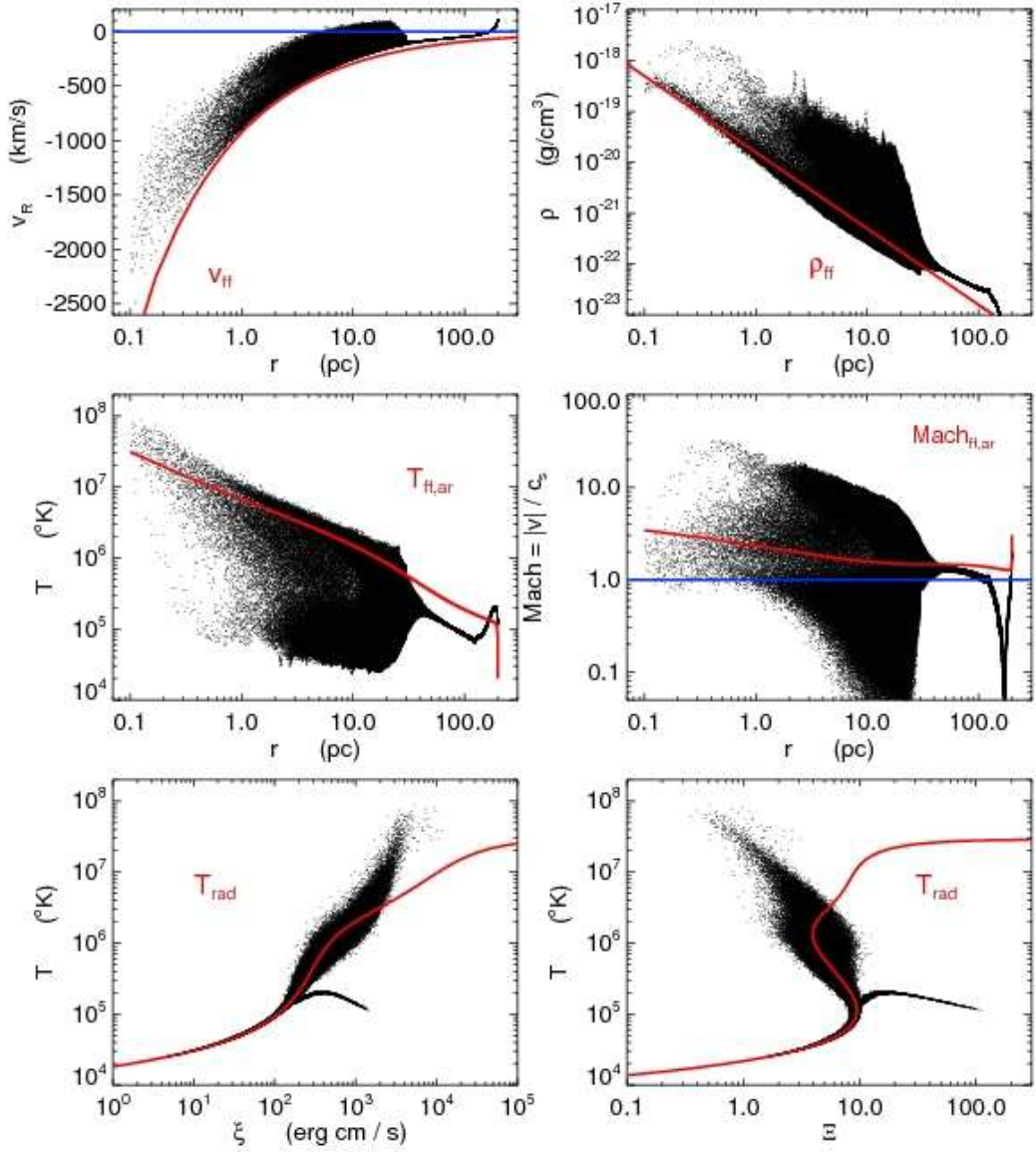}   
\caption{  
Properties of particles in Run $26$ at $t = 2.0$\,Myr   
($0.047$\,Myr earlier than in Figures~\ref{fig-Splash-Lx0.01-30pcYZ} and \ref{fig-Splash-Lx0.01-4pcXY}).   
The top two rows show four quantities vs.~radius: radial velocity, density, temperature and Mach number. 
The bottom row plots the temperature vs.~photo- and pressure-ionization parameters.   
Note that lower photoionization parameter $(\xi = \frac{L_X}{r^2 n})$ corresponds to large radii,   
therefore the points on the left-hand-side in the top two rows are on the right-hand-side in the bottom-left panel.    
The red curve shows the free-fall scaling of the corresponding quantity in the top four panels,  
and the radiative equilibrium scaling in the bottom two panels. 
The blue horizontal straight line in the top-left and middle-right panels indicate $v_r = 0$ and Mach $= 1$. 
}
\label{fig-Scatter-Lx0.01}
\end{figure}

\begin{figure}
\centering
\includegraphics[width = 0.8 \linewidth]{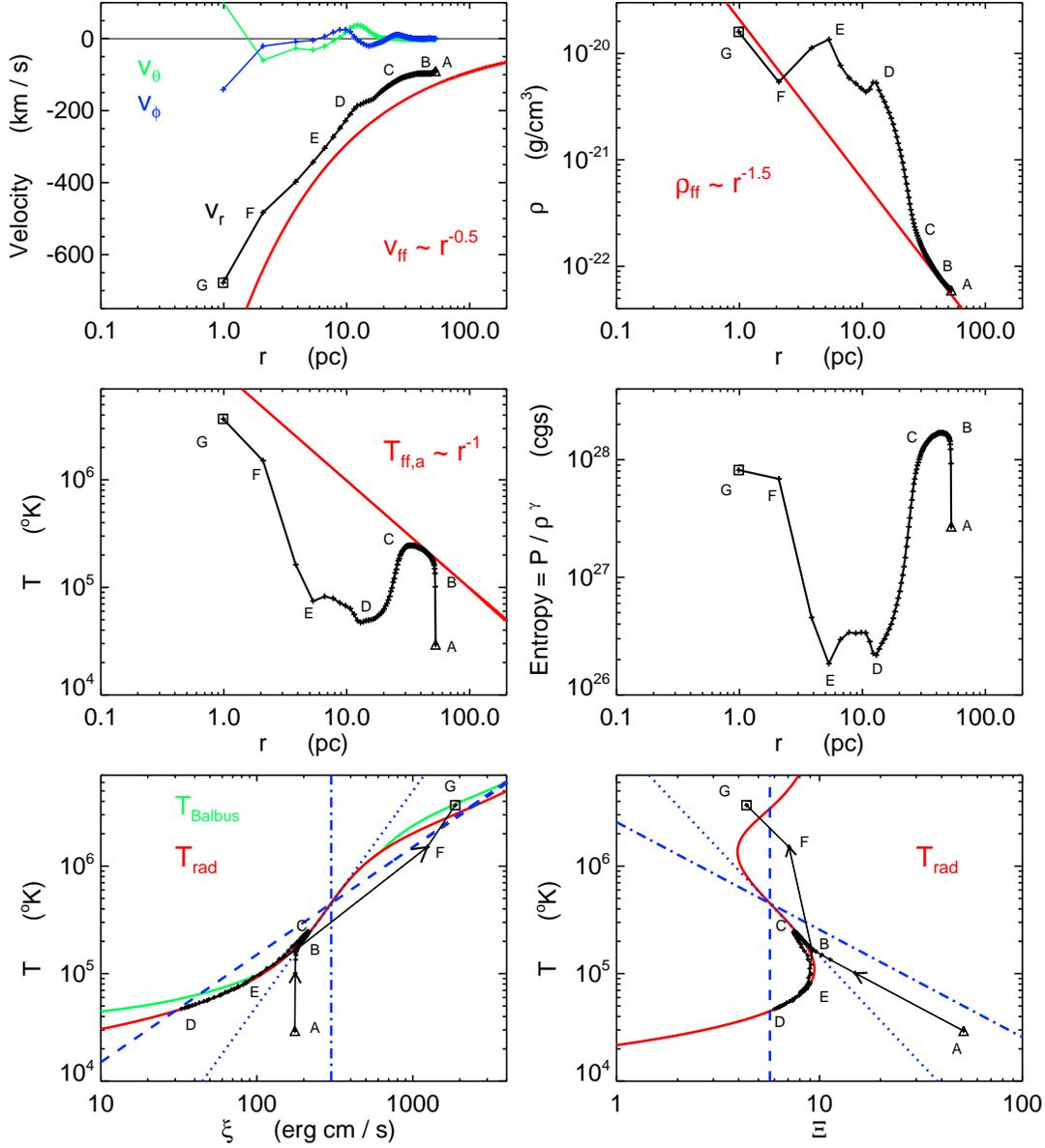}         
\caption{  
Time evolution of a single particle in Run $26$ as it moves inward.  
Letters A, B, C, D, E, F, G denote the major transition points on the track.   
The starting point (A) is denoted by the triangle in each panel, showing 
the initial condition at $t = 1.4$\,Myr and $r = 53$\,pc,  
 The ending point (G) at $t = 1.8$\,Myr and $r = 0.99$\,pc is represented by the square in each panel.  
The plus signs denote the relevant quantity in uniform time intervals of $0.004$\,Myr,   
except the last two points (inner-most in $r$: F, G) which are separated by $\sim 0.001$\,Myr.  
The top two rows show four radial properties: radial and angular velocities, density, temperature and entropy.  
The bottom row plots the evolution in the temperature vs.~photo- and pressure-ionization parameter planes,  
where the direction of progress of time is indicated by arrows.  
The red curve denotes the free-fall scaling of the corresponding quantity in the top two and middle-left panels,  
and the radiative equilibrium scaling in the bottom two panels.  
The green curve in the bottom-left panel is   
the $T - \xi$ solution of the instability criterion from Equation~(5) of \citet{Balbus86}.   
The slopes of three characteristic processes are shown as the blue lines in the bottom row:  
adiabatic free-fall ($T \sim \xi^2$, and $T \sim \Xi^{-2}$) as the dotted line,  
constant-pressure ($T \sim \xi$, and $\Xi =$ constant) as the dashed line, and  
constant-density ($\xi =$ constant, and $T \sim \Xi^{-1}$) as the dash-dotted line.   
These slopes are used to describe various relevant perturbations (see \S\ref{sec-TI} for more details).   
}
\label{fig-Evolution-Single}
\end{figure}

\begin{figure} 
\centering 
\includegraphics[width = 0.6 \linewidth]{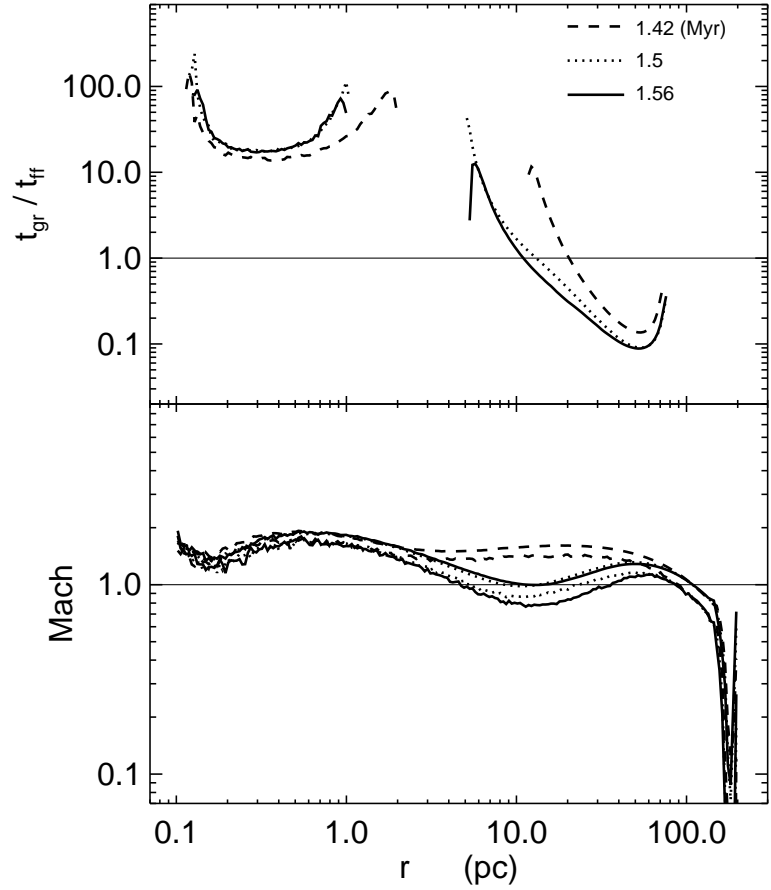}    
\caption{   
Ratio of thermal instability growth time $(t_{gr})$ and free-fall time $(t_{ff})$ in the top panel at three simulation epochs:   
$t = 1.42$\,Myr (dashed curve), $1.5$\,Myr (dotted), and $1.56$\,Myr (solid), when the flow is still spherical.    
The ratio is plotted only at the radial ranges with a positive growth rate and represent regions where TI can grow.   
The radial gaps indicate regions which have negative growth rate and are stable to TI.  
Mach number in the bottom panel; the upper and lower curve of each linestyle denote    
the median and minimum Mach at a given radius.   
Details are described in \S\ref{sec-TI-GrowthTime}.   
}   
\label{fig-timeGrowthTI-Mach}   
\end{figure}

\begin{figure}
\centering
\includegraphics[width = 0.6 \linewidth]{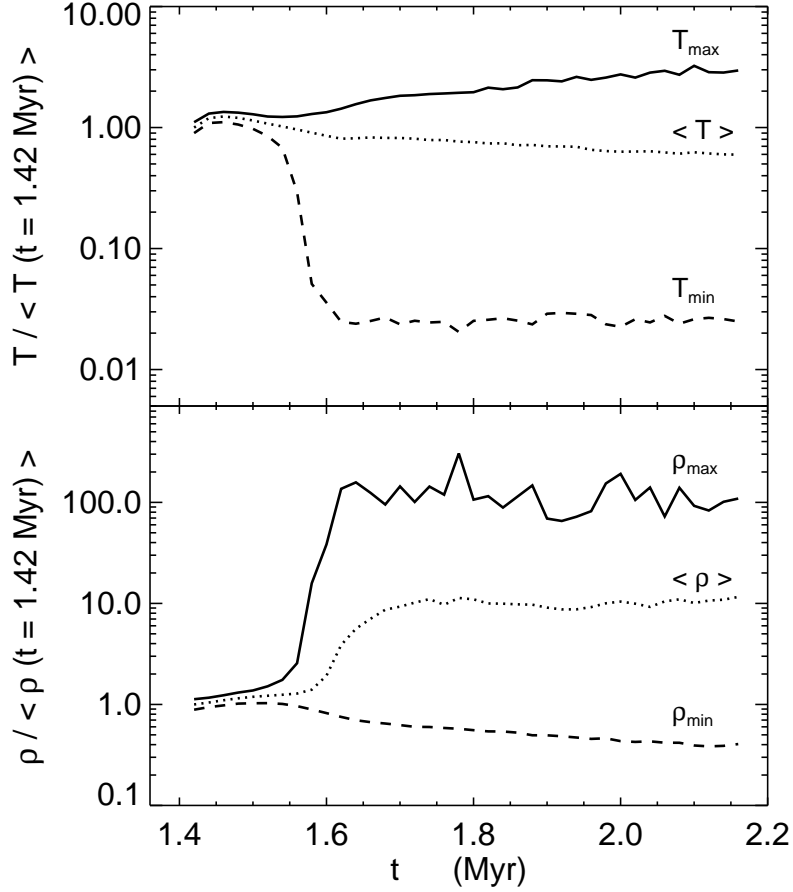}    
\caption{  
Growth of temperature (top panel) and density (bottom panel) perturbation  
caused by TI in Run $26$, at radius $r = 10$\,pc within a width $\Delta r = 1$\,pc, as a function of time.  
Minimum (dashed), maximum (solid), and average (dotted) values  
within the radial bin are plotted as a ratio of the stable unperturbed average value at $t = 1.42$\,Myr.  
The relevant perturbations represented by $T_{\rm min}$, $\rho_{\rm max}$, and $\langle \rho \rangle$    
show an exponential growth (linear on the log scale of the y-axis) for some time initially,    
and then come to a saturation.   
Details are described in \S\ref{sec-TI-GrowthTime}, last two paragraphs.  
}  
\label{fig-Growth-Perturbation}  
\end{figure}

\begin{figure}  
\centering  
\includegraphics[width = 1.0 \linewidth]{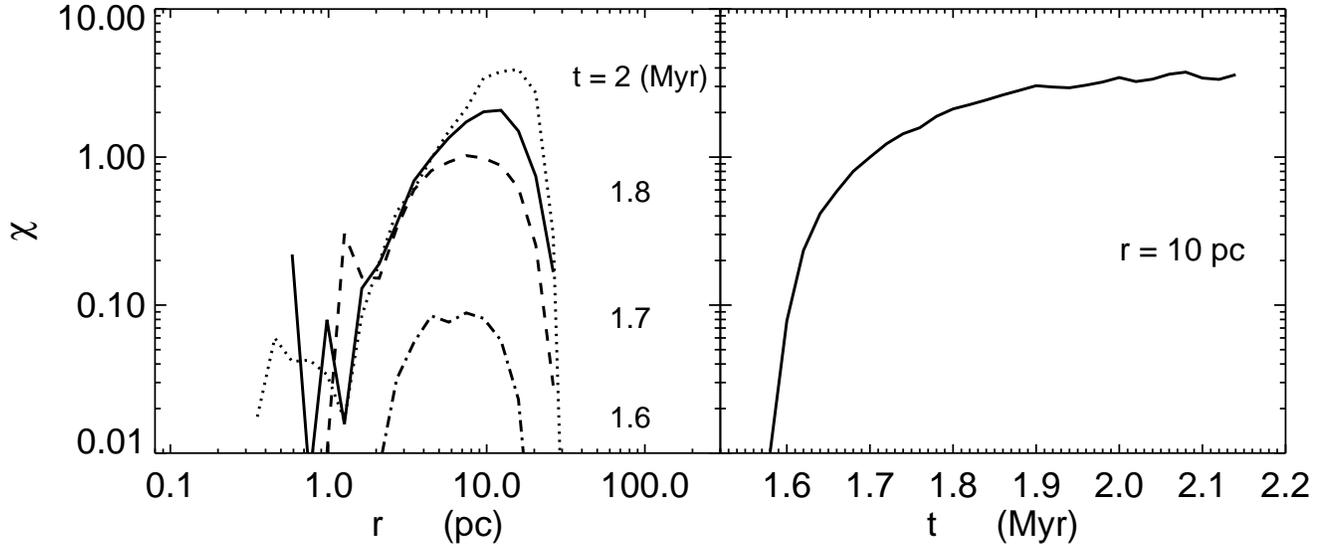}    
\caption{   
Spatial and temporal evolution of the ratio $\chi$  
(defined as the ratio of mass inflow rates of cold phase gas over hot,  
below and above the limiting temperature of $10^5$ K) in Run $26$. 
Left panel shows the radial variation of $\chi$ at four simulation times:  
$t = 1.6$\,Myr (dash-dotted curve), $1.7$\,Myr (dashed), $1.8$\,Myr (solid), and $2.0$\,Myr (dotted).  
$\chi$ is non-zero only within $r \sim 1 - 30$\,pc where there is cold gas.  
Right panel shows the time evolution of $\chi$ at a radius $r = 10$\,pc.  
At a fixed-$r$, $\chi$ increases with time at first, and then becomes saturated at a value $\chi \sim 3 - 4$.  
Details are in \S\ref{sec-Cold-Hot-Frac}.  
}  
\label{fig-alpha_B-Parameter}  
\end{figure}

\begin{figure} 
\centering 
$ 
\begin{array}{cc} 
\includegraphics[width = 0.5 \linewidth]{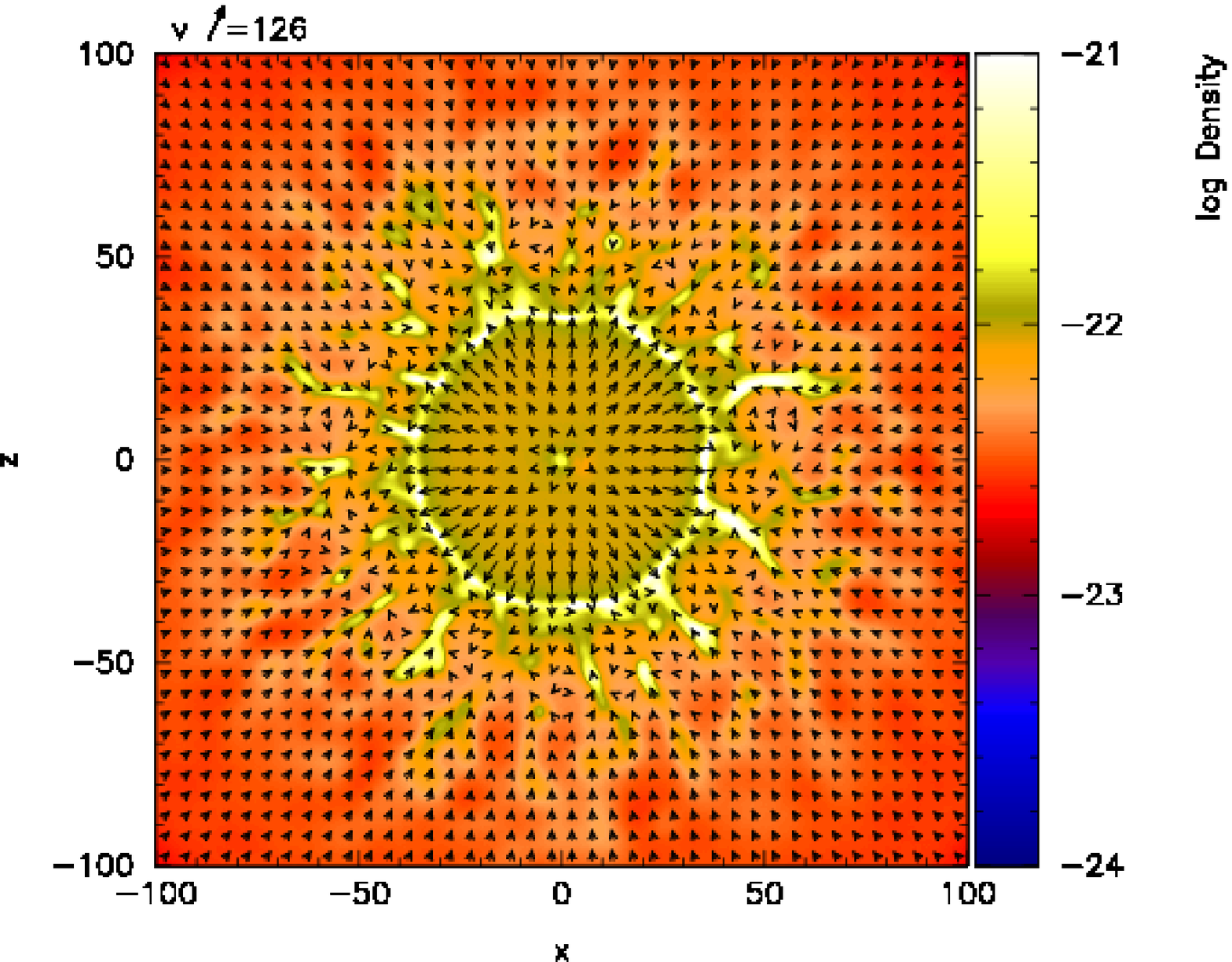} & 
\includegraphics[width = 0.5 \linewidth]{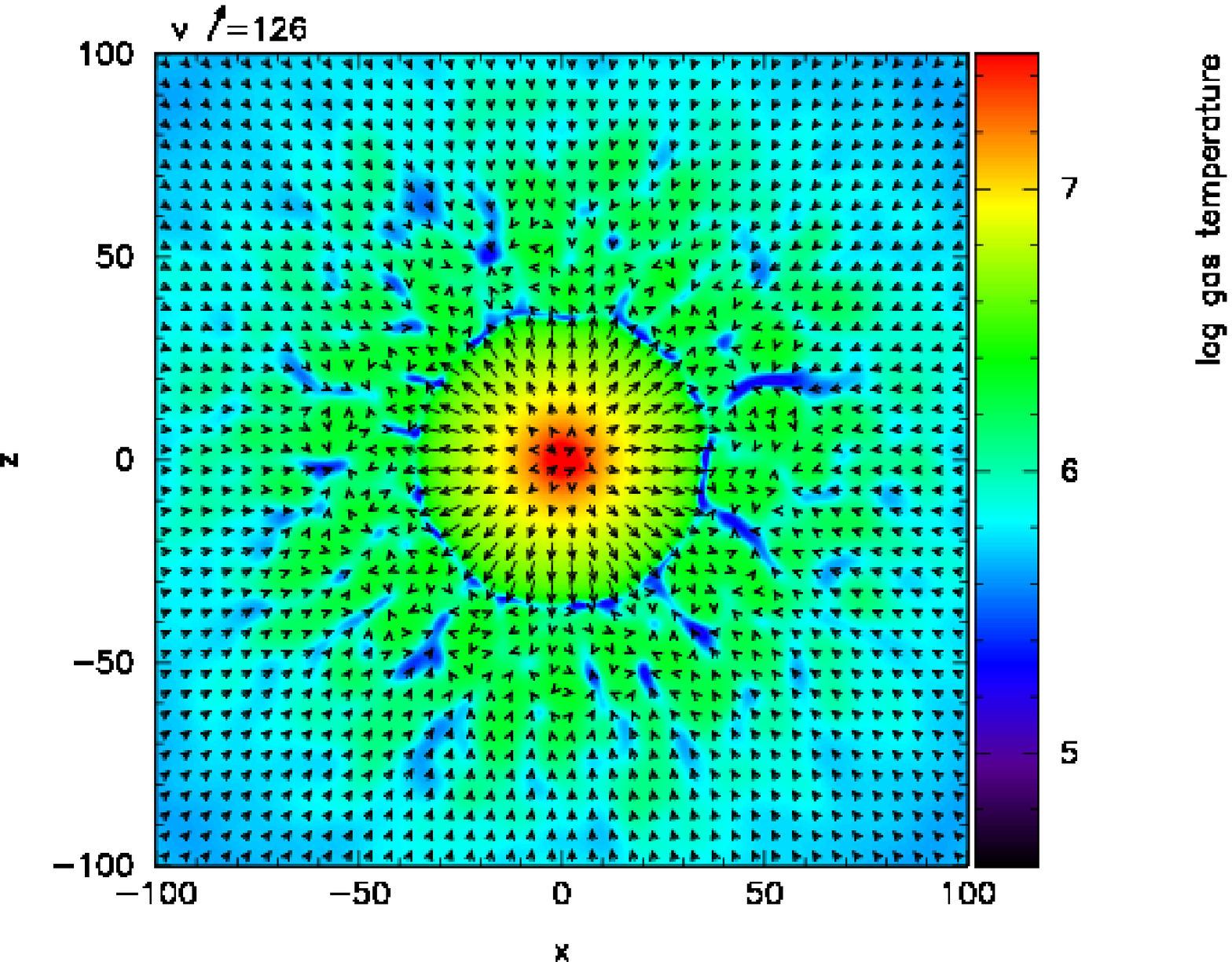} \\ 
\includegraphics[width = 0.5 \linewidth]{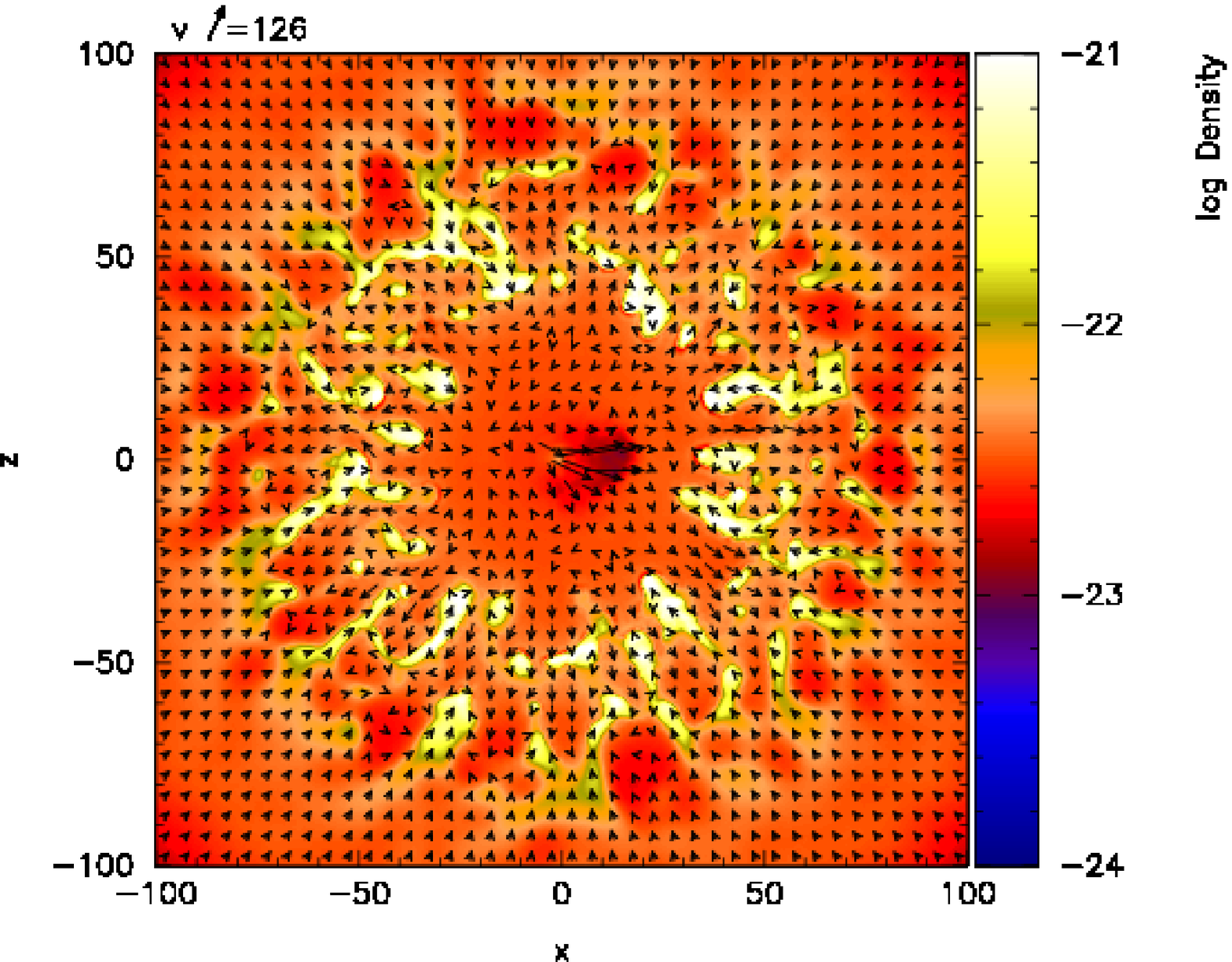} & 
\includegraphics[width = 0.5 \linewidth]{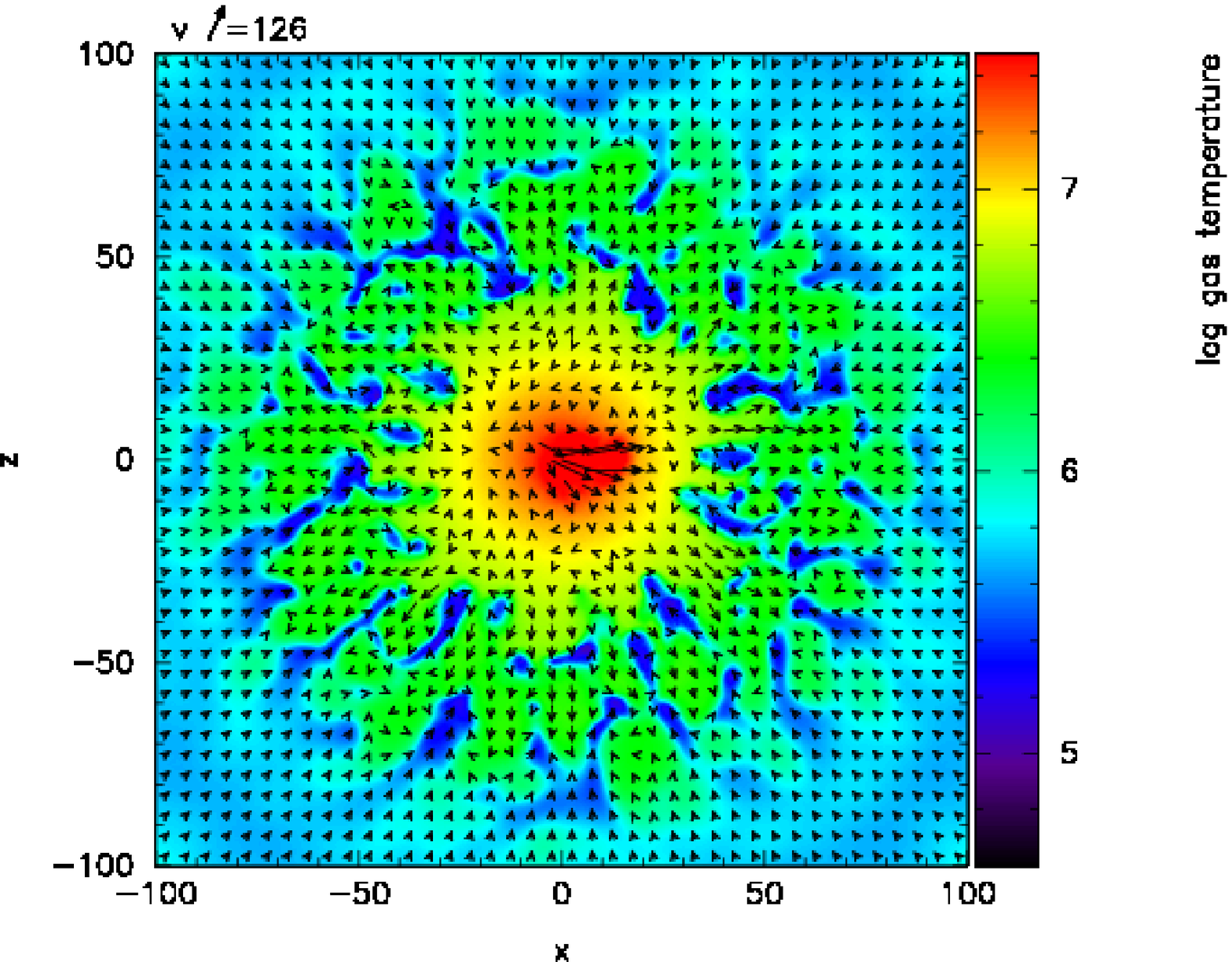} \\ 
\includegraphics[width = 0.5 \linewidth]{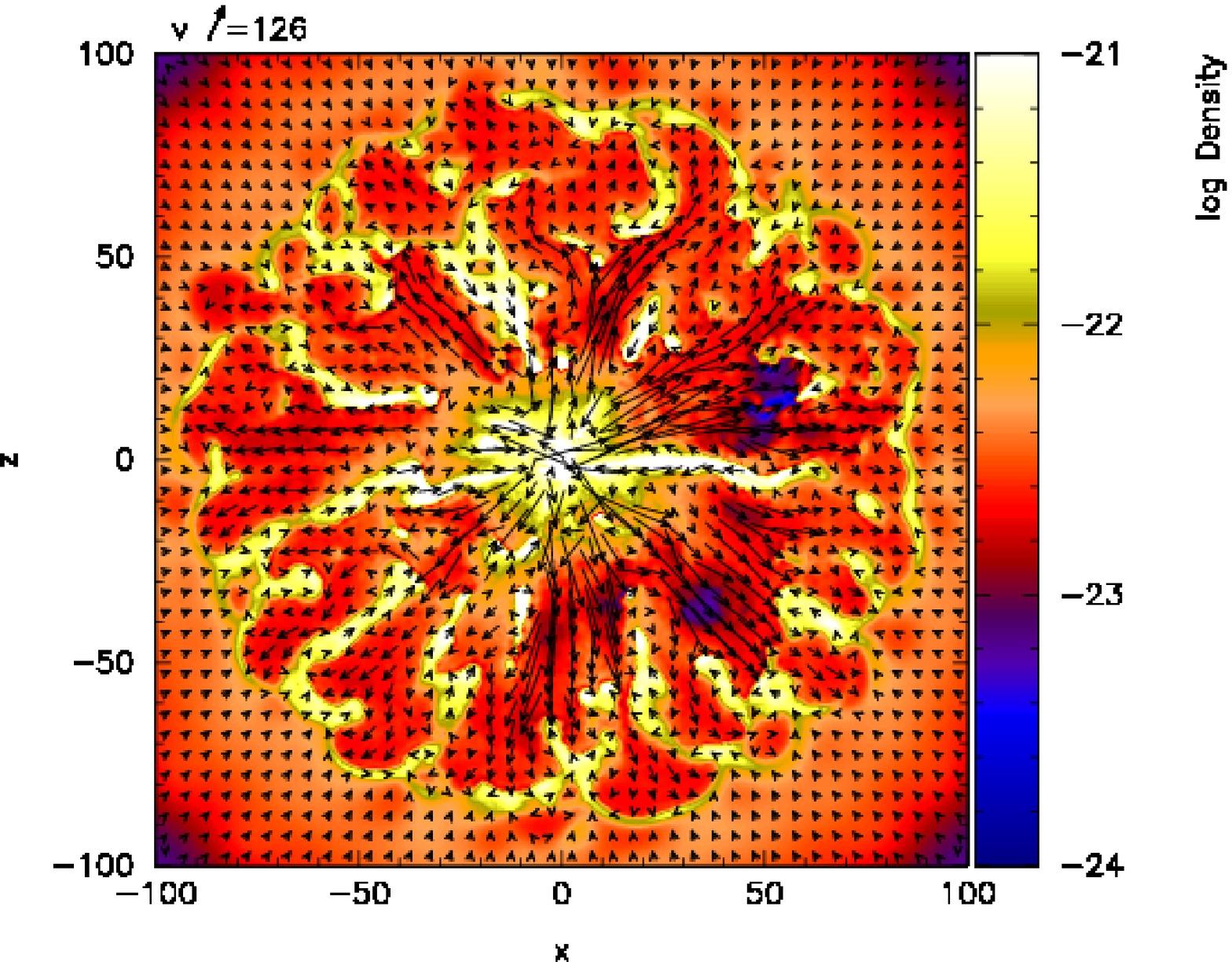} & 
\includegraphics[width = 0.5 \linewidth]{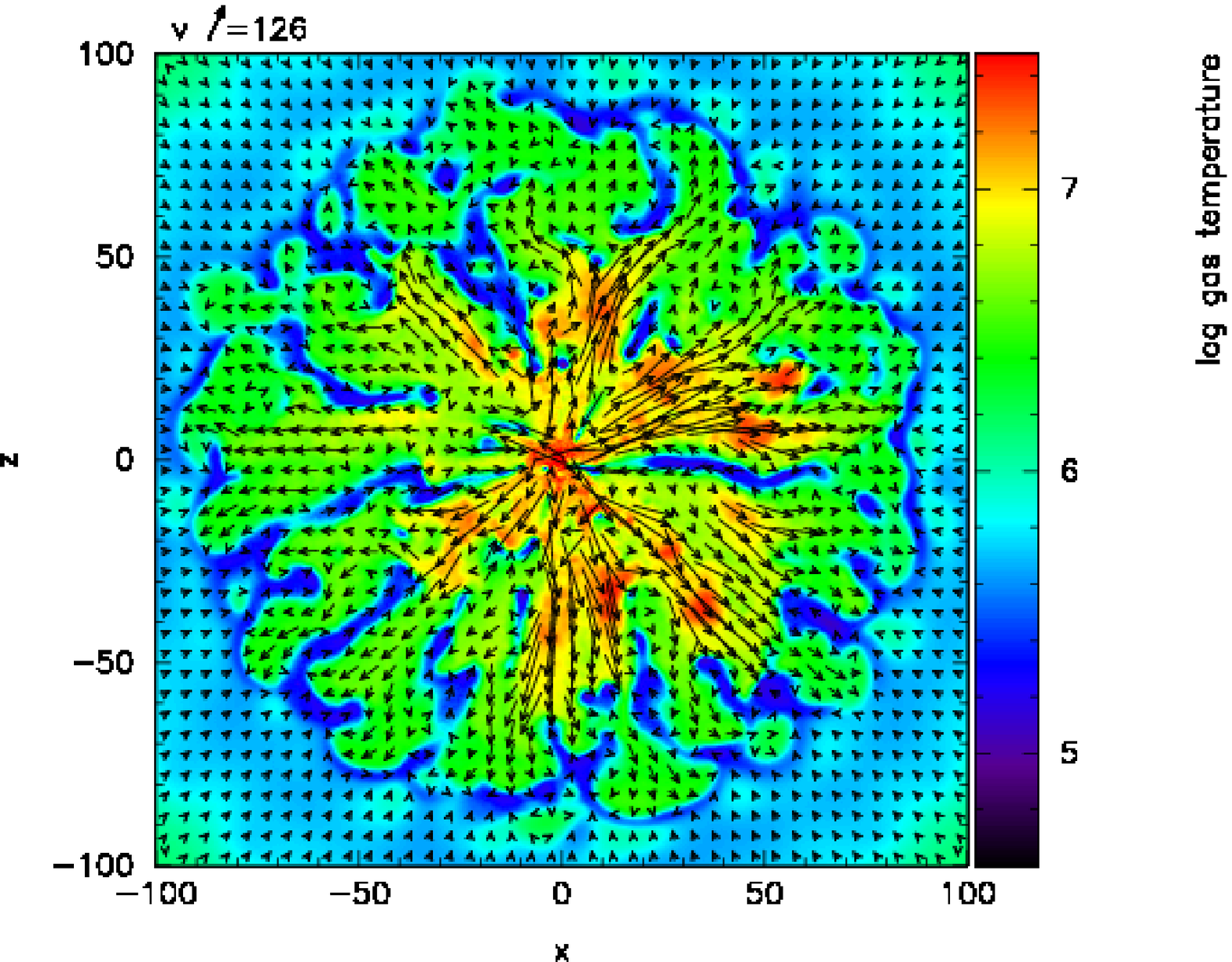} 
\end{array} 
$ 
\caption{ 
Time evolution of gas in Run 27 ($L_{X} / L_{\rm Edd} = 0.02$) showing the   
inner $100$\,pc of the $[x - z]$ plane through $y = 0$.   
The gas density is in the left panel of each row and temperature in the right panel, all in cgs units.    
The velocity vectors are overplotted as arrows,   
whose sizes are denoted at the top-left of each panel in km/s. 
Rows correspond to the times: $t = 1.86$\,Myr (top), $t = 2.12$\,Myr (middle), and $t = 2.46$\,Myr (bottom). 
A central spherically-symmetric outflow inside $r \sim 40$\,pc can be seen in the top row, 
with surrounding gas starting to cool and clump.  
The spherical-symmetry of the outflow is lost in the middle row, 
and the formation of a hot, less-dense bubble can be seen toward the right just off-center. 
The surrounding cool clumps are well-formed and fragmented, and are moving inward. 
The bottom row shows an inhomogeneous mixture of multi-phase gas motion. 
The denser clumps continue falling into the center, but heating up as they move in. 
The hotter gas moves outward, with signatures of few hot bubbles buoyantly rising from the center.   
}
\label{fig-Splash-Lx0.02-100pcXZ}
\end{figure}

\begin{figure} 
\centering 
$ 
\begin{array}{cc}  
\includegraphics[width = 0.5 \linewidth]{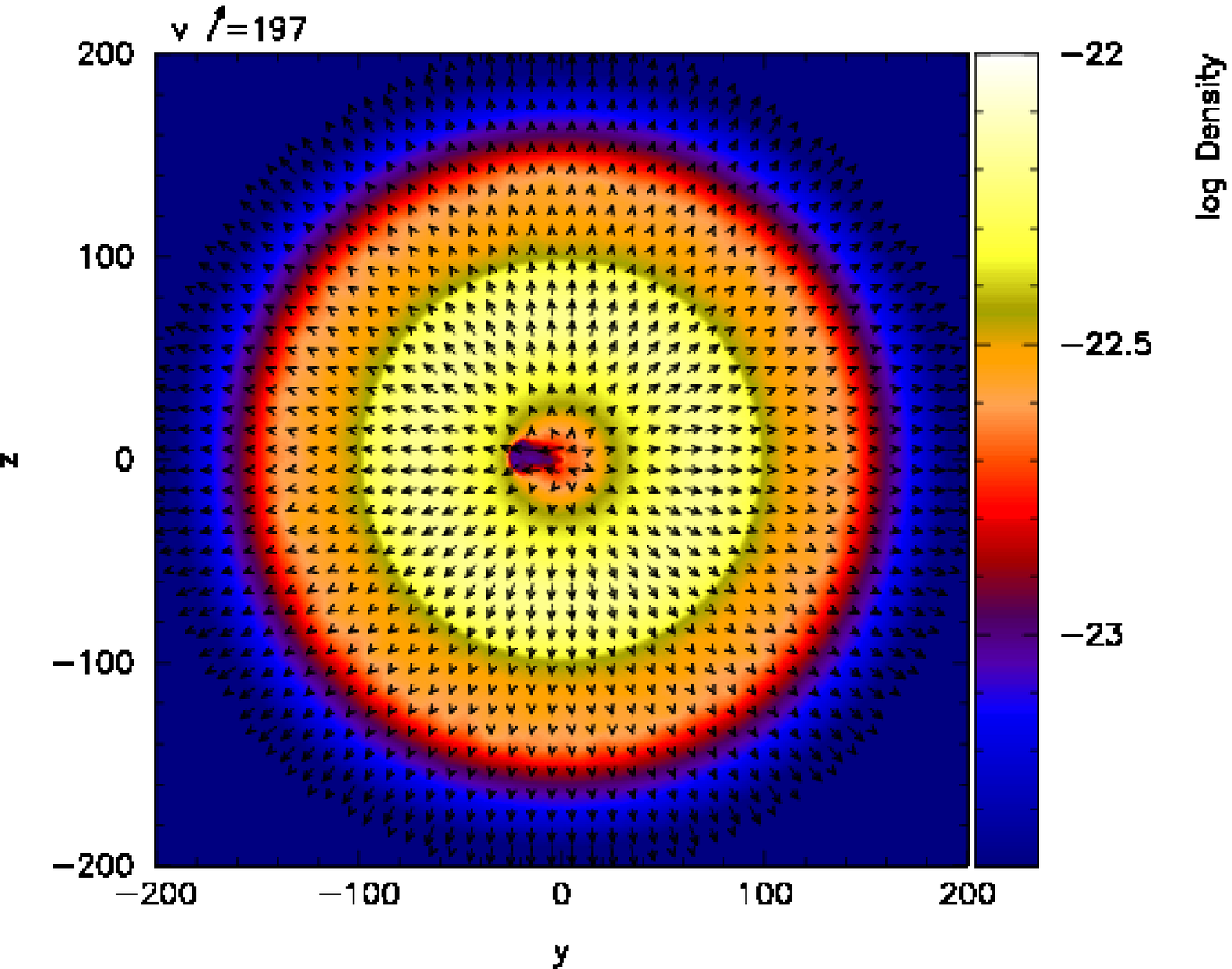} &  
\includegraphics[width = 0.5 \linewidth]{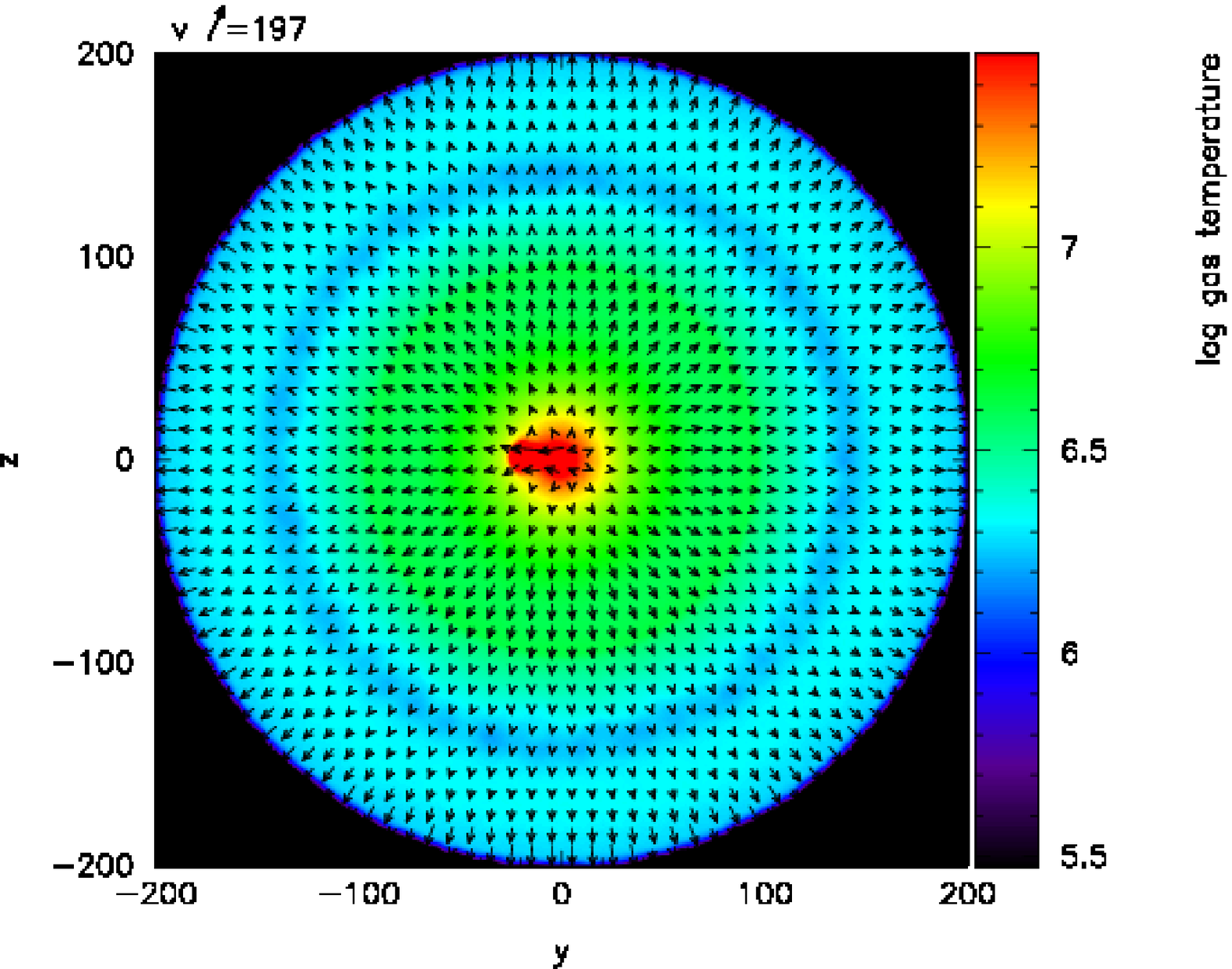} \\  
\includegraphics[width = 0.5 \linewidth]{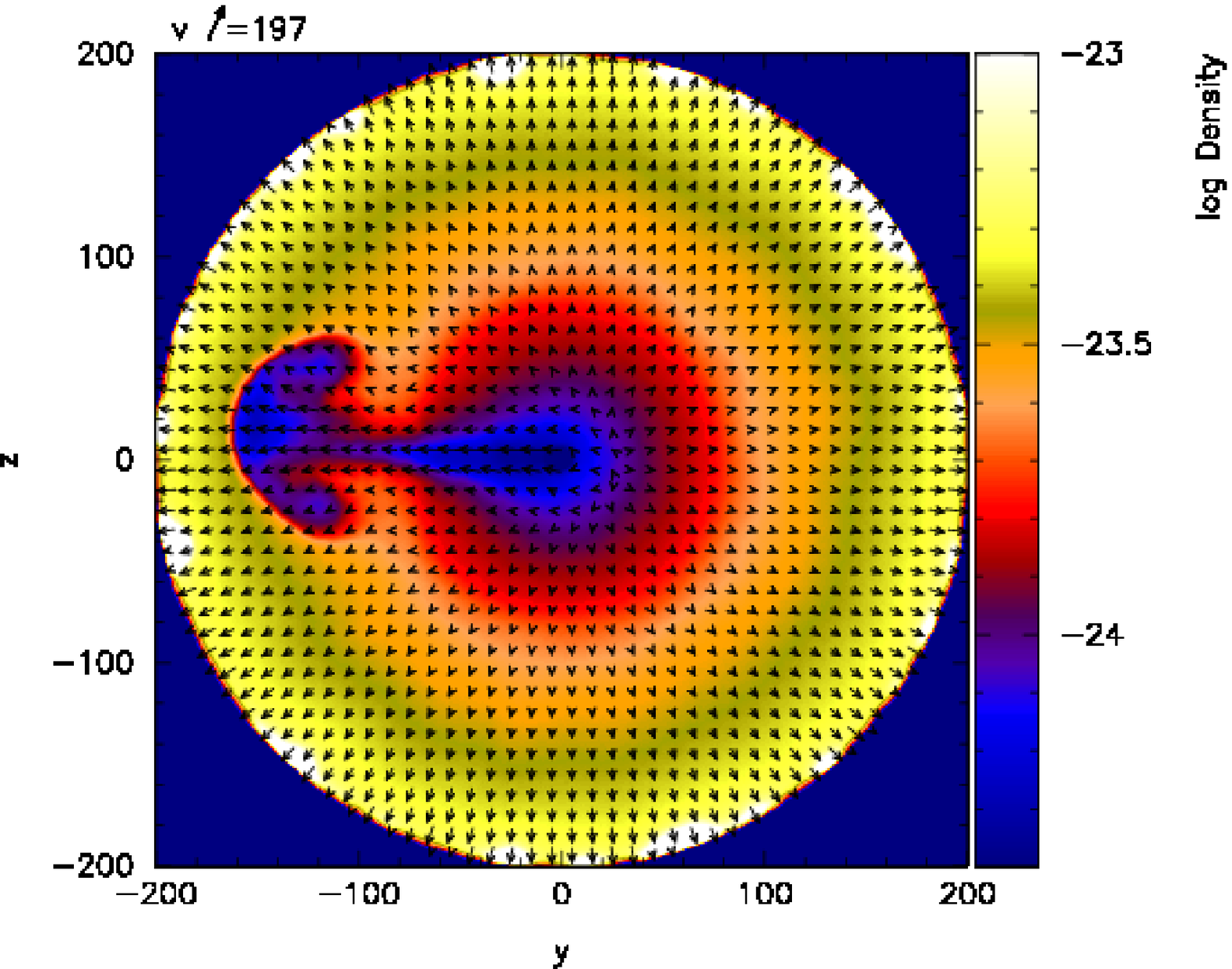} &   
\includegraphics[width = 0.5 \linewidth]{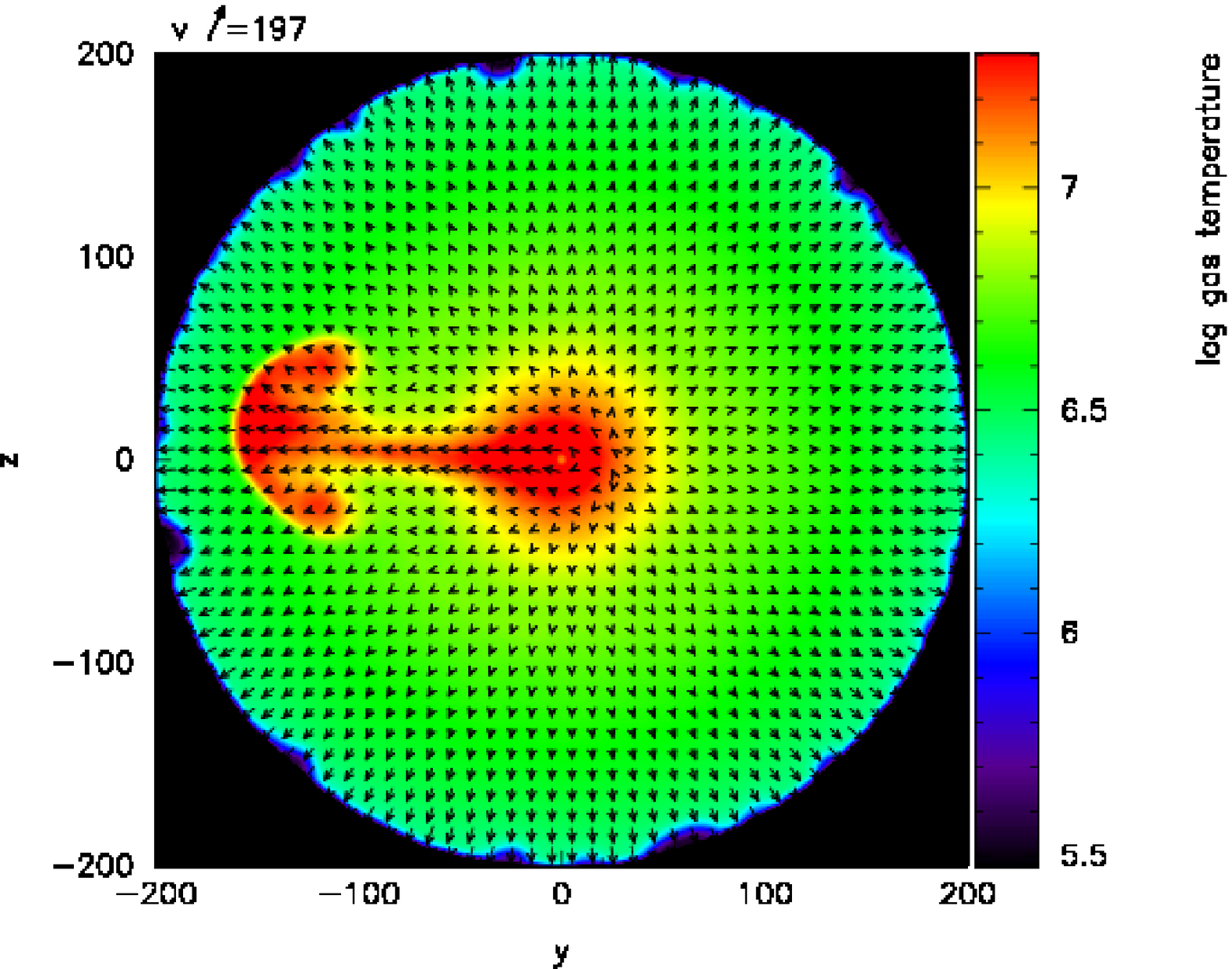}   
\end{array} 
$ 
\caption{ 
Time evolution of gas in Run 28 ($L_{X} / L_{\rm Edd} = 0.05$) showing the whole computational volume 
$200$\,pc of the $[y - z]$ plane through $x = 0$. 
The gas density is in the left panel of each row and temperature in the right panel, overplotted with the velocity vector arrows. 
Rows correspond to the times: $t = 1.8$\,Myr (top), and $t = 3.0$\,Myr (bottom). 
The gas is outflowing in almost all over the volume, except near the very center toward the right in the plotted plane. 
A hot, less-dense bubble is well-formed and buoyantly rises from the center along the negative $z$-axis. 
The rest of the outflowing gas remains spherically symmetric. 
}
\label{fig-Splash-Lx0.05-200pcYZ}
\end{figure}

\begin{figure}
\centering
\includegraphics[width = 0.8 \linewidth]{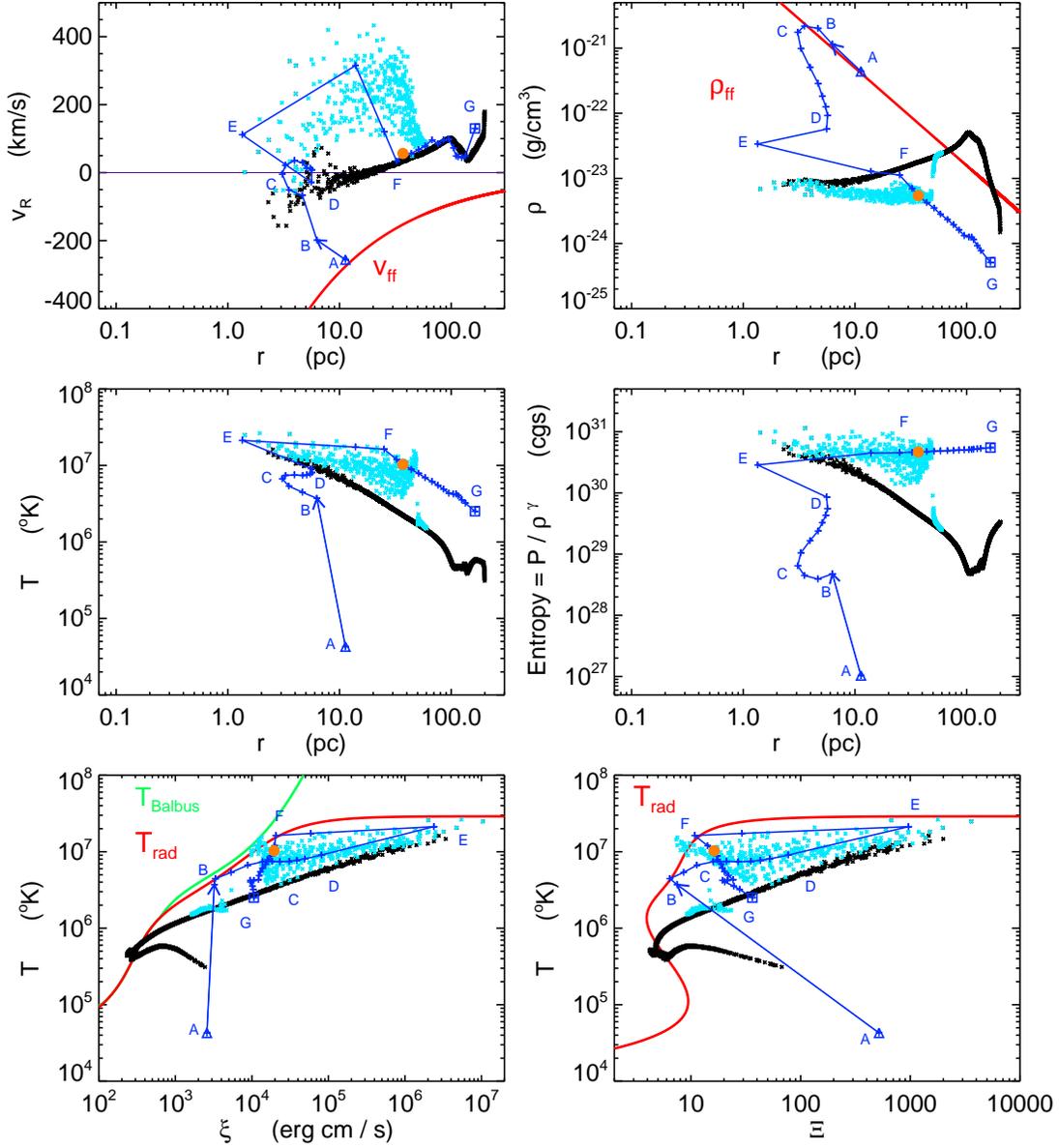}   
\caption{  
Properties of gas in Run 28 at $t = 2.0$\,Myr shown as the cyan and black plus symbols.  
The cyan points are particles overlapping with the negative-$y$ axis (through their smoothing lengths)  
at $r < 60$\,pc hence belonging to the bubble,  
and the black points are particles overlapping with the positive-$z$ axis  
representing rest of the spherically-symmetric volume.  
The top two rows show four quantities vs.~radius: radial velocity, density, temperature and entropy.  
The bottom row plots the temperature vs.~photo- and pressure-ionization parameters.  
The red curve denotes the free-fall scaling of the corresponding quantity in the top two panels,  
and the radiative equilibrium scaling in the bottom two panels.  
The green curve in the bottom-left panel is   
the $T - \xi$ solution of the instability criterion from Equation~(5) of \citet{Balbus86}.  
The blue curve overplotted in each panel represent the time evolution of a single particle inside the bubble,  
as it moves first inward and then outward, in a format similar to Figure~\ref{fig-Evolution-Single}.  
Letters A, B, C, D, E, F, G (written in blue) mark the major transition points of the bubble particle along the track.   
The starting point A from the initial condition at $t = 1.4$\,Myr when it is at $r = 11$\,pc is indicated by the triangle,  
the ending point G at $t = 3.8$\,Myr when it is at $r = 163$\,pc by the square,  
and the direction of time evolution is shown by an arrow in each panel.  
The orange filled circle denotes the particle at the same time $t = 2.0$\,Myr of the snapshot. 
}
\label{fig-Scatter-Lx0.05}
\end{figure}

\begin{figure}  
\centering   
\includegraphics[width = 0.7 \linewidth]{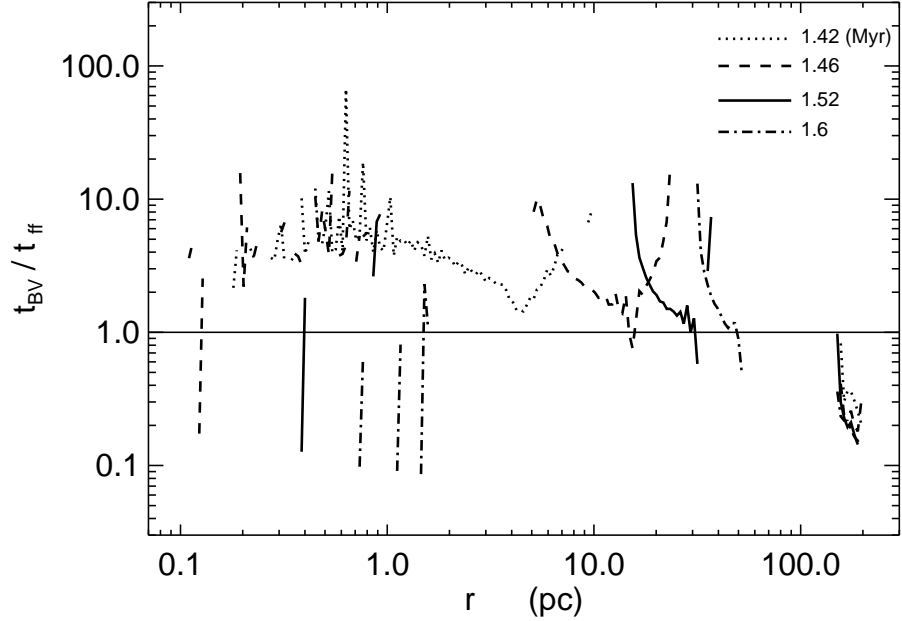}   
\caption{  
Ratio of the Brunt-Vaisala timescale $(t_{BV})$ for convective instability and free-fall time $(t_{ff})$   
at four simulation epochs when the flow is nearly steady:   
$t = 1.42$\,Myr (dotted), $1.46$\,Myr (dashed),  $1.52$\,Myr (solid), and $1.6$\,Myr (dot-dash).   
The ratio is plotted only at the radial ranges having a real effective Brunt-Vaisala frequency    
$(\omega_{BV}^2 > 0)$ and represent convectively unstable regions.      
The radial gaps indicate regions which have imaginary frequency $(\omega_{BV}^2 < 0)$ and are convectively stable.  
The details are described in \S\,\ref{sec-Convection-Time}.  
}  
\label{fig-BruntVaisalaTime-FF}  
\end{figure}

\begin{figure}  
\centering   
\includegraphics[width = 0.7 \linewidth]{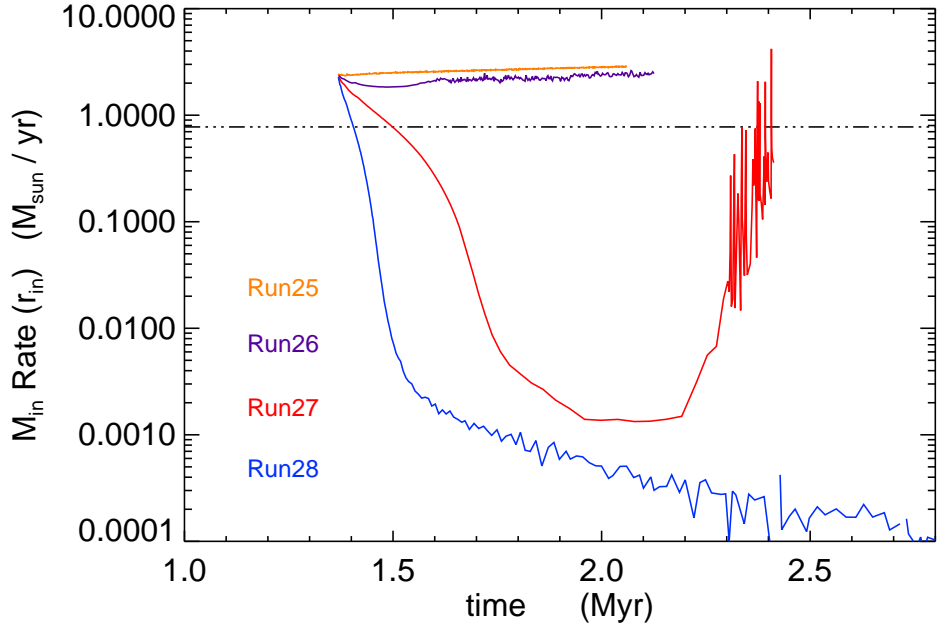}   
\caption{  
Mass inflow rate $\dot{M}_{{\rm in}, r_{\rm in}}$ across the inner boundary as a function of time for the four runs in Table~\ref{Table-AccrModes}, 
with $L_{X} / L_{\rm Edd} = 0.005$ (Run 25), $0.01$ (Run 26), $0.02$ (Run 27), and $0.05$ (Run 28).  
The mass inflow rate decreases as $L_{X}$ is increased, which is associated with the development of an outflow. 
The noisy $\dot{M}_{{\rm in}, r_{\rm in}}$ in Runs 26 and 27 is because of the accretion of multiphase medium, 
with the spikes corresponding to the accretion of cold dense clumps.   
The dash-dotted horizontal line marks the Bondi accretion rate   
corresponding to the $\rho_{\infty}$ and $T_{\infty}$ of the simulations.  
The details are discussed in \S\,\ref{sec-AccretionModes}. 
} 
\label{fig-Mdot_rin} 
\end{figure}

\begin{figure}  
\centering   
$ 
\begin{array}{c}  
\includegraphics[width = 0.6 \linewidth]{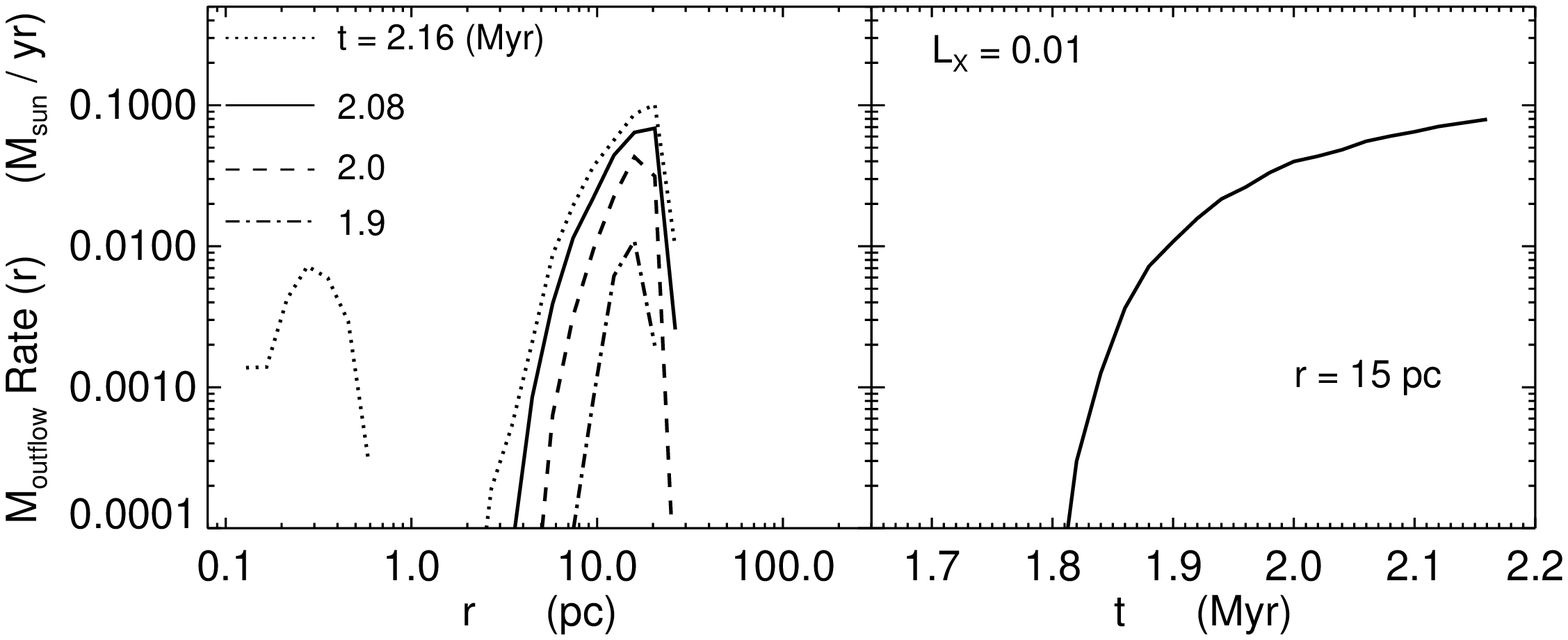} \\  
\includegraphics[width = 0.6 \linewidth]{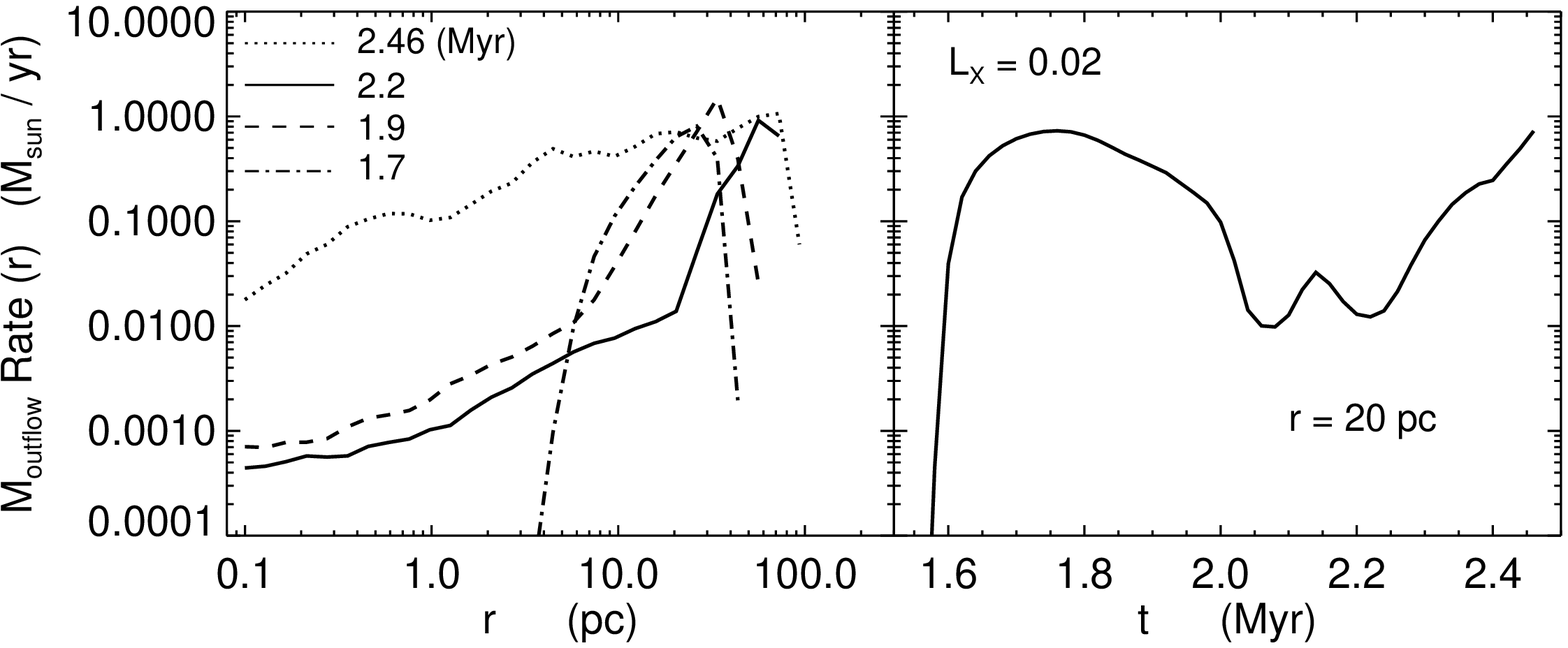} \\  
\includegraphics[width = 0.6 \linewidth]{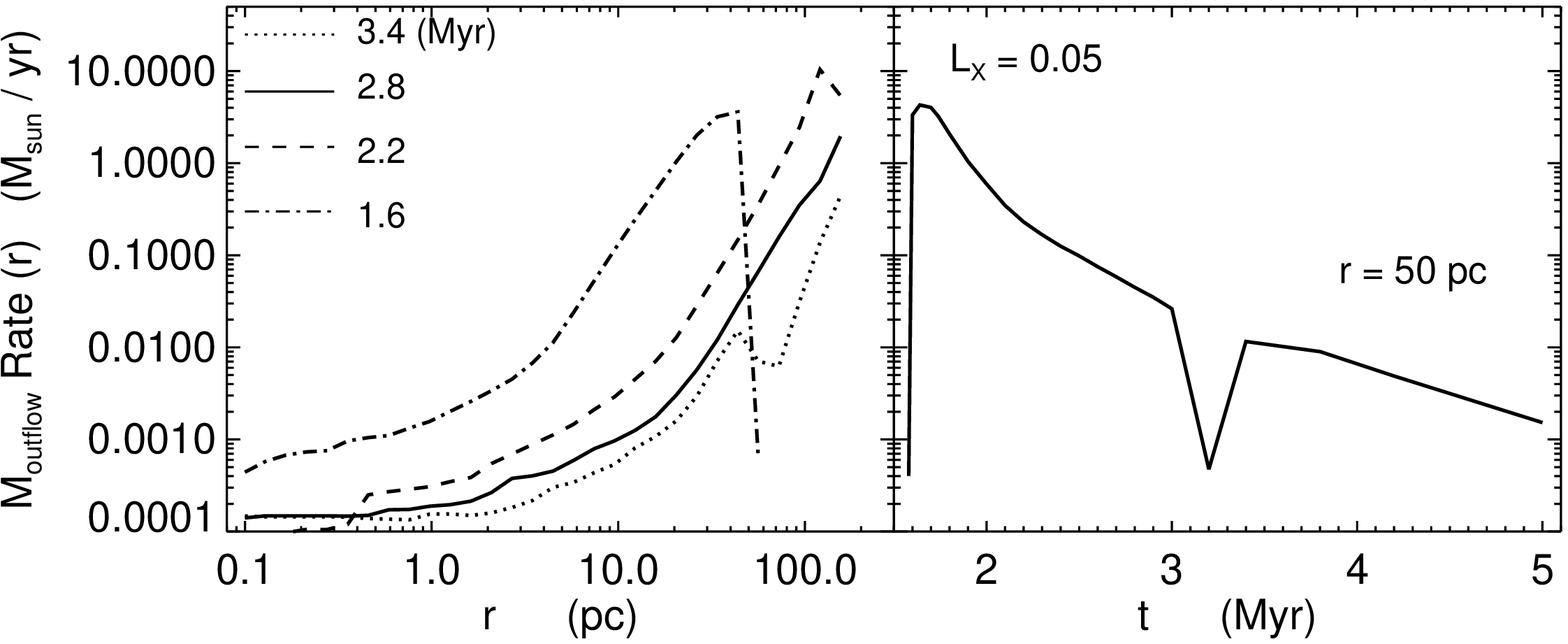}   
\end{array} 
$ 
\caption{   
Mass outflow rate as a function of radius and time for the three runs,   
with $L_{X} = 0.01$ (Run 26, top), $0.02$ (Run 27, middle), and $0.05$ (Run 28, bottom).   
$\dot{M}_{\rm out}$ increases as $L_{X}$ is increased,    
which is associated with the development of strong outflows in runs with higher $L_{X}$.   
The details are discussed in \S\,\ref{sec-AccretionModes}.  
} 
\label{fig-Mdot_Outflow} 
\end{figure}

\end{onecolumn}

\end{document}